\documentclass[preprint]{elsarticle}
\usepackage{setspace}

\usepackage{algorithm}
\usepackage{algorithmic}
\usepackage{multirow}
\usepackage{dcolumn}
\usepackage{amsmath}
\usepackage{flushend}
 \usepackage{cite}
 \usepackage{color}
\usepackage{amssymb}
\usepackage{psfig}
\usepackage{changebar}
\usepackage{graphicx}
\usepackage{subfigure}
\usepackage{multirow}
\usepackage{array}
\usepackage{enumerate}
\usepackage{balance}
\usepackage{cases}
\usepackage{mathrsfs,amssymb}
\usepackage{stfloats}
\usepackage{url}
\usepackage{slashbox}

\newtheorem{theorem}{Theorem}[section]

\newtheorem{proposition}[theorem]{Proposition}

\newenvironment{proof}[1][Proof]{\begin{trivlist}
\item[\hskip \labelsep {\bfseries #1}]}{\end{trivlist}}

\newproof{pf}{Proof}

\usepackage{cont}

\usepackage{amssymb}

\journal{Signal Processing: Image Communication}

\begin{document}

\begin{frontmatter}



\title{Approximate Message Passing-based Compressed Sensing Reconstruction with Generalized Elastic Net Prior}
\tnotetext[t1]{This work was supported by the Natural Sciences and Engineering Research Council (NSERC) of Canada under grant RGPIN312262 and STPGP447223.}


\author{Xing Wang}
\ead{xingw@sfu.ca}
\author[rvt]{Jie Liang\corref{cor1}}
\ead{jiel@sfu.ca}

\cortext[cor1]{Corresponding author}
\fntext[fn1]{The authors are with the School of Engineering Science, Simon Fraser University, Burnaby, BC, Canada.}

\address{
School of Engineering Science, Simon Fraser University, Burnaby, BC, Canada
}

\begin{abstract}
In this paper, we study the compressed sensing reconstruction problem with generalized elastic net prior (GENP), where a sparse signal is sampled via a noisy underdetermined linear observation system, and an additional initial estimation of the signal (the GENP) is available during the reconstruction. We first incorporate the GENP into the LASSO and the approximate message passing (AMP) frameworks, denoted by GENP-LASSO and GENP-AMP respectively. We then focus on GENP-AMP and investigate its parameter selection, state evolution, and noise-sensitivity analysis. A practical parameterless version of the GENP-AMP is also developed, which does not need to know the sparsity of the unknown signal and the variance of the GENP. Simulation results with 1-D data and two different imaging applications are presented to demonstrate the efficiency of the proposed schemes.
\end{abstract}


\end{frontmatter}


\section{Introduction}
\label{sec_intro}

The problem of reconstructing a sparse signal from its noisy linear measurement is crucial to many applications. In this case, the observation $y \in {\mathbb{R}^m}$ can be written as
\begin{equation}
\label{x_model}
y = A x + w,
\end{equation}
where $x \in {\mathbb{R}^n}$ is a $k$-sparse signal, {\it i.e.}, with $k$ nonzero entries ($k \ll n$). $A \in {\mathbb{R}^{m \times n}}$ is a known linear measurement matrix, and $w \in {\mathbb{R}^m}$ is an additive white Gaussian noise with variance $\sigma^2$, {\it i.e.}, $w\sim\mathcal{N}({\mathbf{0}},{\sigma ^2}{\mathbf{I}})$. In this paper, the following ratios are frequently used:
\begin{equation}
\label{kmn_def}
\delta = m/n,  \;\;
\varepsilon = k / n, \;\;
\rho = \varepsilon / \delta = k/m.
\end{equation}
When $m < n$, the problem is underdetermined and has been studied extensively recently via the compressed sensing (CS) theory. It is shown in \citep{RIP} that when $A$ satisfies certain condition and $m$ is larger than some bound, $\ell_1$-based algorithms can successfully recover the sparse signal. Many reconstruction algorithms have been developed to estimate the sparse signal $x$ from $y$, including, {\it e.g.}, convex optimization \citep{RIP}, greedy method \citep{Coherence}, and iterative thresholding algorithm \citep{IST}. However, precise performance analyses of these methods are not available.

Estimation theory can also be used to analyse the performance of CS. In \citep{Replica}, with the help of the replica method from statistical physics, a sharp prediction is derived for the performance of the LASSO or Basis Pursuit Denoising method (BPDN) \citep{lasso,BPDN}, which is an ${\ell_1}$-regularized least-square optimization problem. However, the replica assumption is not rigorous and it cannot be checked for specific problems.

In \citep{AMPoriginal,AMP,Graphical,AMPthesis}, an approximate message passing (AMP) algorithm is developed, which reduces the complexity of classic message passing \citep{Sumproduct}. More importantly, the AMP is rigorous and can predict the final reconstruction performance accurately. Some generalizations of AMP have been developed. For example, in \citep{GAMP}, a generalized AMP (GAMP) is developed to handle arbitrary noise distributions and arbitrary prior distributions. In \citep{EMAMP}, the Gaussian mixture model and expectation-maximization (EM) algorithm are used to learn the distribution of the signal's nonzero coefficients. The AMP also offers a unified framework to exploit other prior knowledge or side information (SI) about the signal \citep{Graphical,AMPthesis}, {\it e.g.}, non-negativity or positivity constraint \citep{AMPthesis,Vila132} and non-uniformly sparsity \citep{StructuredAMP}. Other forms of SI can also be incorporated in the AMP. For example, in \citep{AMP-MMV}, the support of the signal is time-invariant and the signal amplitudes are slowly varying over time. In \citep{DCS-AMP}, the support of the signal is also allowed to change over time.

In this paper, we consider another kind of SI where there is an initial estimation of the sparse signal $x$. Intuitively, this initial estimation can help the reconstruction of $x$. For example, compared to the case without any side information, better reconstruction quality or faster convergence can be achieved with the same sampling rate. This kind of SI could exist in many applications. For example, in dense sensor networks, the sample of a sensor can be estimated from those of its neighboring sensors. This can help the encoding of the sample, as shown in the distributed source coding \citep{XiongDSC}. As another example, in hybrid multiview imaging systems (as demonstrated in Sec. \ref{sec_exp}), some cameras are traditional cameras and some are CS cameras \citep{Trocan01,singlePixel,LiWei,Parmida,XingIcassp13}. Since neighboring cameras are very close to each other, strong correlations exist among their views. Without losing the generality, we assume that the left and right cameras are traditional cameras while the middle camera is a CS camera. Therefore, by exploiting the geometric relationship between neighboring views, disparity estimation and depth-based image rendering techniques can be used to obtain a prediction of the middle view from its neighboring views. As another example, in dynamic systems, the current state can be estimated from the previous state through the state evolution equation \citep{DynamicCS,ModifiedCS,AMP-MMV,DCS-AMP}.

In this paper, we model the initial estimation or SI of the signal as a noisy version of the unknown sparse signal, and modify the LASSO and AMP frameworks to incorporate the initial estimation. In \citep{Hui}, an additional ${\ell_2}$ penalty term is added to LASSO, and the scheme is called elastic net-regularized LASSO. In the optimization framework derived in Sec. \ref{sec_inf} of this paper, there is also an additional $\ell_2$ penalty term to LASSO. When the SI is zero, our scheme reduces to that in \citep{Hui}. Therefore the SI in our framework can be viewed as a generalized elastic net prior (GENP), and we denote the GENP-aided LASSO and AMP as GENP-LASSO and GENP-AMP respectively. Although \citep{SZ13} is the first to study elastic net prior using AMP, it focuses on the binary classification problem and there is no theoretical performance analysis.

After developing the frameworks of GENP-LASSO and GENP-AMP, we focus on the GENP-AMP, and investigate its parameter selection, state evolution, asymptotic prediction performance, and noise-sensitivity analysis. However, these theoretical analyses require the knowledge of the sparsity of the unknown sparse signal and the variance of the generalized elastic net prior. In practices, these parameters have to be estimated. In \citep{Paraless}, a parameterless AMP is developed using Stein's unbiased risk estimate (SURE). Inspired by \citep{Paraless}, we apply the SURE theory to GENP-AMP and develop a parameterless version of GENP-AMP. Simulation results with 1-D data and two different imaging applications are presented to demonstrate the efficiency of the proposed schemes.

\subsection{Related work}

There have been some efforts on exploiting various initial estimations in CS. One example is the CS problem with partially known support \citep{ModifiedCS}, which shows that by finding the signal that satisfies the measurement constraint and is the sparsest outside the partially known support, the CS reconstruction can be improved, and bounds on the reconstruction error are derived. However, the method is time-consuming. Another relevant approach is to recover the estimation error instead of the sparse signal \citep{Trocan01}, based on the assumption that the prediction error between the initial estimation and the sparse signal is sparser than the signal itself, and is thus easier to be recovered, but this method lacks theoretical analysis. It is also possible that the prediction error is denser than the original sparse signal, if the initial estimation has poor quality.

In \citep{Parmida}, the belief-propagation-based CS framework (BPCS) in \citep{CSBP} is used to exploit the SI from neighboring cameras in multiview imaging systems, where the SI is used as the starting point for belief propagation. In \citep{XingIcassp13}, a squared-error-constrained penalty term is added to the CS of multiview images. It also considers a more general case, where the variances of the prediction errors are different at different entries. A fast solution is developed based on the Gradient Projection for Sparse Reconstruction (GPSR) algorithm \citep{GPSR}.

The sparsity-constrained dynamic system estimation scheme proposed in \citep{DynamicCS} and the dynamic compressed sensing via approximate message passing (DCS-AMP) proposed in \citep{AMP-MMV,DCS-AMP} are closely related to our framework. In \citep{DynamicCS}, a prediction of the signal is obtained from the state evolution model, and the norm of the prediction error is added as a penalty term in the objective function of LASSO or BPDN method. In \citep{AMP-MMV,DCS-AMP}, the sparse signal is modeled as the Bernoulli-Gaussian distribution and the correlation between the active amplitudes in different time slots is assumed to be a stationary steady-state Gaussian-Markov process. The EM and AMP are applied to learn the hidden parameters and perform the inference. Although the model in \citep{AMP-MMV,DCS-AMP} is similar to ours, it relies on sequential data to learn the hidden parameters, and cannot be applied to solve the problem discussed here directly. In fact, it is not clear how to extend the method in \citep{AMP-MMV} to solve the problem in this paper.

Several papers have also studied the theoretical contribution of the prior knowledge \citep{ModifiedCS,Kamilov}. In \citep{ModifiedCS}, the authors have provided some sharp bounds on the necessary number of CS measurements to successfully reconstruct the original sparse signal, based on nullspace property and geometry interpretations. However, it is mainly on the noiseless case. The performance of noisy case remains unknown. Kamilov \emph{et al}. have taken the first step towards a theoretical understanding of EM-based algorithms \citep{AMP-MMV,DCS-AMP,Kamilov}, although the complete analysis is still not available. Our method does not involve any loose constant, and can accurately predict the performance.

On the other hand, the GENP considered in this paper can be incorporated into the GAMP \citep{GAMP}. However, even if the GENP is known to the GAMP, the GAMP still needs to know the exact prior distribution $p(x)$. Therefore in practice some learning-based methods such as the EM algorithm have to be used to learn $p(x)$ \citep{EMAMP}. Our scheme does not need to know $p(x)$, and only assumes that $x$ is sparse.

In Sec. \ref{sec_exp} of this paper, we will present simulation results with 1-D data and two different imaging applications. We will show that the overall performance of our methods is better than the AMP, GAMP, the method in \citep{Trocan01}, the modified CS in \citep{ModifiedCS}, the linear minimum mean squared error (LMMSE) method, and direct denoising. Our parameterless method also works very well below the phase transition boundary of AMP, although its performance still needs to be improved above the boundary, because the estimated variance of the prior using the method in \citep{Paraless} is unstable in this case.

Some preliminary results of this paper have been reported in \citep{XingIcassp14}. Due to the importance of the problem of side/prior information-based CS reconstruction, earlier versions of this paper have received attention from other researchers  \citep{Mota14,Renna14}. In \citep{Mota14}, only the noiseless CS sampling scenario is considered, and an $\ell_1$ or $\ell_2$ constraint of the prior information is added to the $\ell_1$ Basis Pursuit objective function of the unknown signal. However, only some loose bounds of different constraints are presented in it.  In \citep{Renna14}, the classification and reconstruction of high-dimensional signals from low-dimensional features in the presence of side information is discussed. The high-dimensional signals are assumed to follow Gaussian mixture model (GMM) that can be learned from training data. The fundamental limits are derived based on this assumption. In our paper, we do not make any assumption about the target signal except sparsity and there is no training data involved.

\section{Background: Minimax MSE of Soft Thresholding Algorithm}
\label{sec_bkgd}

In this section, we briefly review the minimax MSE of the soft thresholding algorithm \citep{AMP,softshrink}, which plays an important role in AMP. Suppose we need to recover a $k$-sparse $n$-vector ${x^0} = ({x^0}(i):  i \in [n])$ (where $[n] \equiv \{1, \ldots, n\}$) contaminated by a Gaussian white noise, {\it i.e.},
\begin{equation}\nonumber
y(i) = x^0(i) + z^0(i), \; i \in [n],
\end{equation}
where ${z^0}(i)\sim \mathcal{N}(0,{\sigma ^2})$ is independent and identically distributed. One way to estimate the signal is to solve the following LASSO or $\ell_1$-regularized least-squares problem,
\begin{equation}
\label{eq_lasso}
\hat{x} ^\lambda = \mathop {\arg \min }\limits_x {\text{ }}\frac{1}
{2}\left\| {y - x} \right\|_2^2 + \lambda {\left\| x \right\|_1}.
\end{equation}
An important fact is that the solution of this problem is equivalent to that of the well-known soft thresholding algorithm in wavelet denoising \citep{softshrink},
\begin{equation}\nonumber
 {\hat{x} ^\lambda }(i) = \eta (y(i);\lambda ), \; i \in[n],
\end{equation}
where the soft thresholding operation with threshold $\theta $ is
\begin{equation}
\label{threshold}
\eta (x;\theta ) =
\begin{cases}
  x - \theta & \mbox{if } x > \theta, \\
  0 & \mbox{if } -\theta  \leqslant x \leqslant \theta,   \\
  x + \theta & \mbox{if } x <  - \theta.
\end{cases}
\end{equation}

A reasonable choice of the threshold $\lambda$ in (\ref{eq_lasso}) is a scaled version of the noise standard deviation,  {\it i.e.}, $\lambda  = \alpha \sigma$. The MSE of the soft thresholding algorithm can thus be written as
\begin{equation}
\label{eta_mse}
\mbox{mse}({\sigma ^2};p,\alpha ) \equiv E\{ {[\eta (X + \sigma Z;\alpha \sigma ) - X]^2}\},
\end{equation}
where the expectation is with respect to independent random variables $Z\sim \mathcal{N}(0,1)$ and $X\sim p$.

The soft thresholding method is scale-invariant \citep{AMP}, {\it i.e.},
\begin{equation}
{\text{mse}}({\sigma ^2};p,\alpha ) = {\sigma ^2}{\text{mse}}(1;{p_{1/\sigma }},\alpha ),
\end{equation}
where ${p_s}$ is a scaled version of $p$, ${\text{ }}{p_s}(S) = p(\{ x:{\text{ }}s x \in S\})$. Therefore we only need to focus on $\sigma  = 1$, and the notation ${\text{mse(}}1;p,\alpha )$ can be simplified into ${\text{mse(}}p,\alpha )$.

Since ${x^0}$ is $k$-sparse, we can define the following set of probability measures with small non-zero probability,
\begin{equation}
\label{F_ep}
{\mathcal{F}_\varepsilon } \equiv \{ {p}:{\text{ }}{p}{\text{ is a probability measure with }}{p}(\{ 0\} ) \geqslant 1 - \varepsilon \} ,
\end{equation}
where $\varepsilon=k/n$ is defined in (\ref{kmn_def}).

The minimax threshold MSE is thus defined as \citep{AMP}
\begin{equation}
\label{minimaxmse}
{M^ \pm }(\varepsilon ) = \mathop {\inf }\limits_{\alpha  > 0} \mathop {\sup }\limits_{p \in {\mathcal{F}_\varepsilon }} {\text{mse}}(p,\alpha ),
\end{equation}
which is the minimal MSE of the worst distribution in $\mathcal{F}_\varepsilon$, where $\pm$ means a nonzero estimand can take either sign.

For a given $\alpha$, the worst case MSE in (\ref{minimaxmse}) is given by \citep{AMP}
\begin{equation}
\label{supmse}
\mathop {\sup }\limits_{p \in {F_\varepsilon }} {\text{mse}}(p,\alpha ) = \varepsilon (1 + {\alpha ^2}) + (1 - \varepsilon )[2(1 + {\alpha ^2})\Phi ( - \alpha ) - 2\alpha \phi (\alpha )],
\end{equation}
with $\phi (z) = \exp ( - {z^2}/2)/\sqrt {2\pi } $ being the standard normal density, and $\Phi (z) = \int_{ - \infty }^z {\phi (x) dx} $ the Gaussian cumulative distribution function. Moreover, the supremum can be achieved by the following three-point probability distribution on the extended real line $\mathbb{R} \cup \{  - \infty ,\infty \} $
\begin{equation}\nonumber
p_\varepsilon ^* = (1 - \varepsilon ){\delta _0} + \frac{\varepsilon }
{2}{\delta _\infty } + \frac{\varepsilon }
{2}{\delta _{ - \infty }},
\end{equation}
where $\delta_t$ is a Dirac delta function at $t$. In practice, we are more interested in the near-worse-case signals with finite values. It is known that the following $c$-least-favorable distribution can achieve a MSE that is a fraction of $(1-c)$ of the worst case,
\begin{equation}
\label{least_favor_p}
{p_{\varepsilon ,c}} = (1 - \varepsilon ){\delta _0} + \frac{\varepsilon }
{2}{\delta _{{h^ \pm }(\varepsilon ,c)}} + \frac{\varepsilon }
{2}{\delta _{ - {h^ \pm }(\varepsilon ,c)}},
\end{equation}
where ${h^ \pm }(\varepsilon ,c) \sim \sqrt{2 \text{log}(\varepsilon^{-1})}$ as $\varepsilon \rightarrow 0$.

\section{GENP-aided LASSO}
\label{sec_inf}

In this paper, we study the generalized elastic net prior (GENP)-aided CS reconstruction, where in addition to the CS sampling as in (\ref{x_model}), an initial estimation of $x$, denoted by $\tilde x$, is available during reconstruction, which can be seen as a noisy version of $x$. The error of this estimation, $e = \tilde x - x$, is assumed to be i.i.d. additive white Gaussian with variance $\sigma_s^2$, {\it i.e.}, $e \sim  \mathcal{N} ({\mathbf{0}}, \sigma_s^2{\mathbf{I}})$. This Gaussian noise model is decently accurate in applications such as image acquisition with poor illumination, high temperature, or transmission error, and has been widely used in image denoising \citep{BM3D}. The ratio between the noise variance of $\tilde{x}$ and that of the compressed sampling noise in Eq. (\ref{x_model}) will be used later for noise sensitivity analysis.
\begin{equation}
\gamma _s^2 = \sigma _s^2/{\sigma ^2}.
\label{gamma}
\end{equation}

To exploit the $\tilde{x}$ in the CS reconstruction, we propose the following optimization formula,
\begin{equation}
\label{minimization}
\begin{split}
 \hat x(\lambda ,{\tau _s}) = \mathop {\arg \min }\limits_{z \in {\mathbb{R}^n}} & \left( \frac{1} {2}\left\| {y - Az} \right\|_2^2 \right. \\
& \left. + \lambda {\left\| z \right\|_1} + \frac{\tau_s }{2}\left\| {\tilde x - z} \right\|_2^2 \right),\\
\end{split}
\end{equation}
which is a generalized version of the LASSO in (\ref{eq_lasso}) with an additional $\ell_2$ penalty term caused by the initial estimation $\tilde{x}$ to ensure the solution close to this initial estimation. When $\tilde{x}=0$, the problem reduces to the elastic net-regularized LASSO in \citep{Hui}. Therefore we call $\tilde{x}$ generalized elastic net prior (GENP), and the problem in Eq. (\ref{minimization}) generalized elastic net prior-aided LASSO (GENP-LASSO).

A special case of our framework is that when $p(x)$ follows the Laplacian distribution, the result of Eq. (\ref{minimization}) is equivalent to the maximum a posteriori (MAP) solution. However, our framework in Eq. (\ref{minimization}) is more general than this special case because we do not rely on any assumption about $p(x)$, except that $x$ should be sparse as defined in Eq. (\ref{F_ep}). In the following theoretical analysis, we will apply the minimax estimator introduced in Sec. \ref{sec_bkgd} to study the parameter selection, state evolution and MSE performance of the optimization problem in Eq. (\ref{minimization}).

Similarly, although the GENP in our framework can also be incorporated into the GAMP scheme in \citep{GAMP}, it should be noted that GAMP also needs to know the exact prior distribution $p(x)$. Therefore learning algorithms such as the EM have to be used to learn the prior distribution \citep{EMAMP}. In Sec. \ref{sec_exp}, we will compare our method to the EMGMAMP in \citep{EMAMP} and a modified EMGMAMP that incorporates the GENP, and show that our method has better overall performance.

In LASSO, the ratio $\rho $ in Eq. (\ref{kmn_def}) cannot be larger than 1, {\it i.e.}, the number of selected atoms is bounded by the number of samples, whereas it is shown in \citep{Hui} that in the elastic net-regularized LASSO, the quadratic penalty term removes this limitation. Our noise sensitivity analysis in Sec. \ref{noise_sensitivity} will show that $\rho<1$ is also not necessary in the GENP-LASSO.

The parameters $\lambda$ and ${\tau _s}$ in Eq. (\ref{minimization}) are closely related to $\sigma _s^2$, the noise variance of the GENP. How to tune the two parameters $\lambda$ and ${\tau _s}$ will be addressed later in the paper. 

The proposed GENP-LASSO in (\ref{minimization}) is a convex optimization problem and can be solved by, {\it e.g.}, the interior point methods (as used in the CVX package \citep{cvx}) and the gradient methods. For example, to incorporate the GENP into the Orthant-Wise Limited-memory Quasi-Newton (OWLQN) algorithm \citep{OWLQN}, which is a popular gradient-based method for large-scale LASSO problems, we can replace the ${\ell _2}$ regularization term $\left\| z \right\|_2^2$ in it by the quadratic penalty term $\left\| {\tilde x - z} \right\|_2^2$. However, both interior point and gradient methods are quite slow for large-scale problems.

In this paper, we will solve the GENP-LASSO problem by modifying the fast AMP algorithm, which enjoys several advantages, {\it e.g.}, low complexity and the capability of predicting the final performance accurately.

Note that we can also combine $y$ and $\tilde{x}$ as follows.
\begin{equation}
\label{eq_ls}
\left[ \begin{gathered}
  y \hfill \\
  {\tilde x} \hfill \\
\end{gathered}  \right] = \left[ \begin{gathered}
  A \hfill \\
  I \hfill \\
\end{gathered}  \right]x + \left[ \begin{gathered}
  w \hfill \\
  e \hfill \\
\end{gathered}  \right].
\end{equation}
This is an overdetermined system of $x$ with $(m + n)$ equations. Therefore $x$ can be solved directly using the least-squares (LS) or the linear minimum mean squared error (LMMSE) method. However, we will show in Sec. \ref{sec_exp} that the performance of the LMMSE method is not as good as the proposed method (the LS solution is even worse than that of the LMMSE, and is not included due to space limitation). Note that the LMMSE solution also requires the knowledge of $p(x)$.

\section{GENP-aided Approximate Message Passing}
\label{sec_siamp}

In this section, we present the formulae of GENP-AMP. We then study its connections with the GENP-LASSO, and derive its corresponding parameter selections and state evolution.

\subsection{The Formula of GENP-AMP}

In \citep{Graphical}, the following iterative formulas of AMP are obtained after simplifying the traditional min-sum-based message passing algorithm using the quadratic approximation.
\begin{equation}
\label{amp_01}
\begin{gathered}
  \hat x_0^t = {x^t} + {A^T}{r^t}, \hfill \\
  {x^{t + 1}} = \eta (\hat x_0^t;{\theta _t}), \hfill \\
 \end{gathered}
\end{equation}
\begin{equation}
{b_t} = \frac{1}{m}{\left\| {{x^t}} \right\|_0},
\end{equation}
\begin{equation}
\label{amp_02}
{r^t} = y - A{x^t} + {b_t}{r^{t - 1}}.
\end{equation}

Each iteration of AMP only needs to update the estimate $x^t$ in (\ref{amp_01}) and the residual $r^t$ in (\ref{amp_02}), which have only $m+n$ entries. The complexity is thus much lower than traditional message passing methods that need $2mn$ updates. Note that the AMP is parameterized by two sequences of scalar parameters: the thresholds $\{{\theta _t}\}_{t \ge 0}$ and the factors $\{b_t\}_{t \ge 0}$.

To incorporate the GENP into AMP, we modify the local message of each AMP variable node from $\lambda {\left\| z \right\|_1}$ to $\lambda {\left\| z \right\|_1} + \frac{{{\tau _s}}}{2}\left\| {\tilde x - z} \right\|_2^2$. By the same simplifications and derivations in \citep{Graphical}, we can get the following iterative estimate of the $n$-vector signal $x$. The details are skipped due to space limitation.
\begin{equation}
\label{amp_x}
\hat x_0^t = \frac{{{u_t}}}{{1 + {u_t}}}\tilde x + \frac{1}{{1 + {u_t}}}({x^t} + {A^T}{r^t}),
\end{equation}
\begin{equation}
{\text{ }}{x^{t + 1}} = \eta (\hat x_0^t;\;{\theta _t}),
\end{equation}
\begin{equation}
\label{b_t}
 {b_t} = \frac{1}
{{1 + {u_{t - 1}}}}\frac{{{{\left\| {{x^t}} \right\|}_0}}}
{m},
\end{equation}
\begin{equation}
\label{amp_r}
r^t = y - A x^t + b_t  r^{t-1}.
\end{equation}
Compared to AMP, $\hat{x}_0^{t}$ in our scheme is a linear combination of $x^t + A^Tr^t$ and the GENP, adaptively controlled by a new sequence of scalar parameters, $\{u_t\}_{t \ge 0}$. The factor $b_t$ is also affected by $u_{t-1}$. When $u_t=0$, $\tilde{x}$ has no contribution, and the proposed framework reduces to the standard AMP in \citep{AMPoriginal,AMP,Graphical, AMPthesis}. The iteration is applied to each entry. Hence, if the variances of different ${{\tilde x}_i}$ are different, the method can still be applied by changing the scalar ${u_t}$ to vector $\mathbf{{u_t}} = [{u_{t,1}},{u_{t,2}},...,{u_{t,n}}]$ and the scalar ${\theta _t}$ to its vector case.

\subsection{Connections to GENP-LASSO}

As shown in \citep{Graphical}, the parameters ${\{ {\theta _t}\} _{t \geqslant 0}}$ and ${\{ {b_t}\} _{t \geqslant 0}}$ are constrained by its connection with the min-sum algorithm. This is also true for the new parameter ${\{ {u _t}\} _{t \geqslant 0}}$. However, the following proposition shows that GENP-AMP provides a very general solution for the GENP-LASSO problem in Eq. (\ref{minimization}). When there is no GENP ($u_t=0$), the proposition reduces to Prop. 5.1 in \citep{Graphical} for LASSO.

\begin{proposition}
\label{parameter}
Let $({x^*},{r^*})$ be the fixed point of the GENP-AMP algorithm given by (\ref{amp_x}) and (\ref{amp_r}) for fixed ${\theta _t} = \theta$, ${\text{ }}{u _t} = u$, and ${\text{ }}{b_t} = b$. Then ${x^*}$ is also a minimum of the GENP-LASSO problem in (\ref{minimization}) with
\begin{equation}
\label{lambda}
  \lambda  = (1 + u )\theta (1 - b),
\end{equation}
\begin{equation}
\label{tau}
  \tau_s  = u(1 - b).
\end{equation}
\end{proposition}

\begin{pf}
The fixed-point condition of Eq. (\ref{amp_x}) is
\begin{equation}
{x^*} = \frac{u}
{{1 + u}}\tilde x + \frac{1}
{{1 + u }}({x^*} + {A^T}{r^*}) -  \theta {v^*},
\end{equation}
where $v_i^* = sign(x_i^*)$ if $x_i^* \ne 0$ and $v_i^* \in [ - 1, + 1]$ otherwise. Similarly, from (\ref{amp_r}), we get $(1 - b){r^*} = y - A{x^*}$, or $r^*=(y-Ax^*)/(1-b)$. Plugging into the equation above, we get
\begin{equation}\nonumber
(1+u) \theta (1-b) {v^*} + u(1-b) (x^* - \tilde{x}) = A^T (y - A x^*).
\end{equation}

On the other hand, in Eq. (\ref{minimization}), by setting the derivative of the GENP-LASSO objective function with respect to $z$ to zero, we get the stationary condition
\begin{equation}
\label{stationary}
\lambda {v^*} + \tau_s ({x^*} - \tilde x) = {A^T}(y - A{x^*}).
\end{equation}
Comparing the two equations above leads to the conclusion.
\end{pf}

\subsection{GENP-AMP State Evolution and Parameter Selection}
\label{subsec_state}

In this part, we derive the state evolution of GENP-AMP and investigate its parameter selection. The state evolution was first developed to describe the asymptotic limit of the AMP estimates as $m, n \to \infty$ for any fixed $t$, but with the same sample ratio $\delta=m/n$, as defined in (\ref{kmn_def}) \citep{Graphical}. It enables the accurate prediction of the MSE of AMP by solving a fixed-point equation. This part is based on Sec. IV of \citep{AMP}.

First, we define the MSE map $\Psi $ as
\begin{equation}\nonumber
\Psi ({q^2},u,\delta ,\sigma ,{\sigma_s},\alpha , p ) \equiv \mbox{mse}( \mbox{npi}({q^2},u;\delta ,\sigma ,{\sigma_s});p,\alpha ),
\end{equation}
which is the MSE of the soft thresholding as defined in (\ref{eta_mse}) with ${\text{npi}}$ (noise-plus interference) as the noise variance, where $q^2$ is the variance of the thresholded estimator, and $\text{npi}$ is the variance of the un-thresholded estimator in (\ref{amp_x}), which can be written as (see Appendix A for the derivation)
\begin{equation}
\label{npi}
{\text{npi(}}{q^2},u;\delta ,\sigma ,{\sigma_s}) = {(\frac{u}
{{1 + u}})^2}\sigma_s^2 + {(\frac{1}
{{1 + u}})^2}({\sigma ^2} + \frac{q^2}{\delta }).
\end{equation}

As pointed out in \citep{Graphical}, the choice of the AMP parameter $\theta_t$ can be quite flexible. A good option is ${\theta _t} = \alpha {\xi _t}$, where $\alpha > 0$, and ${\xi _t}$ is the root MSE of the un-thresholded estimation $\hat{x}_0^t$ in (\ref{amp_x}). From this, based on the i.i.d. normalized distribution of $A$ and the large system limit \citep{AMP}, it can be shown that
\beq
\label{xit2}
\xi _t^2 = {\text{npi}}(q_t^2,u_t^2;\delta ,\sigma ,{\sigma _s}) \approx {\left( {\frac{{{u_t}}}
{{1 + {u_t}}}} \right)^2}\sigma _s^2 + {\left( {\frac{1}
{{1 + {u_t}}}} \right)^2}\frac{{\left\| {{r^t}} \right\|_2^2}}
{m}.
\eeq

Besides,  we have ${\left\| {{x^t}} \right\|_0}/n \approx {\text{E}}\{ {{\eta}'}({x_0} + {\sigma ^t}Z;\alpha {\sigma ^t})\}$. According to Eq. (\ref{b_t}, \ref{lambda}, \ref{tau}),  Prop. \ref{parameter} can be rewritten as
\beq
\label{parameter_prediction}
\begin{gathered}
  \lambda  = (1 + {u_*})\alpha {\xi _*}\left[ {1 - \frac{1}
{{1 + {u_*}}}\frac{{{\text{E}}\{ {{\eta}'}({x_0} + {\xi _*}Z;\alpha {\xi _*})\} }}
{\delta }} \right], \hfill \\
  {\tau _s} = {u_*}\left[ {1 - \frac{1}
{{1 + {u_*}}}\frac{{{\text{E}}\{ {{\eta}'}({x_0} + {\xi _*}Z;\alpha {\xi _*})\} }}
{\delta }} \right], \hfill \\
\end{gathered}
\eeq
where
 ${\xi _*} = {\lim _{t \to \infty }}{\xi _t}$. Since the computation of ${q^2}$ is nontrivial,
 Eq. (\ref{xit2}) is useful for practical algorithm design, whereas Eq. (\ref{npi}) is mainly for theoretical analysis.

The state of GENP-AMP is defined as a 7-tuple $(q^2,u;\delta ,\sigma ,{\sigma_s},\alpha ,p)$. The state evolution follows the rule
\begin{equation}\nonumber
 \begin{split}
  (q_t^2,{u_t};\delta ,\sigma ,{\sigma_s},\alpha ,p) &\mapsto (\Psi (q_t^2,{u_t}),{\Upsilon}(q_t^2,{u_t});\delta ,\sigma ,{\sigma_s},\alpha ,p), \\
  t & \mapsto t + 1,
\end{split}
\end{equation}
where $q_t^2$ and ${u_t}$ are the MSE and the weighting parameter in the $t$-th iteration, and $\Psi$ and ${\text{ }}{\Upsilon}$ are the evolution functions of $q_t^2$ and ${u_t}$, respectively.
As $(\delta ,\sigma ,{\sigma_s},\alpha ,\upsilon )$ are fixed during the evolution, we only need the following state evolutions of  $q_t^2$ and ${u_t}$ (See Appendix \ref{apx_state} for the derivation).
\begin{equation}
\label{evolution}
\begin{gathered}
  q_t^2 \mapsto q_{t + 1}^2 \equiv \Psi (q_t^2,\frac{{{\sigma ^2} + q_t^2/\delta }}{{\sigma _s^2}}),{\text{ }} \hfill \\
  {u_t} \mapsto {u_{t + 1}} = {\Upsilon}(q_t^2,{u_t}) = \frac{{{\sigma ^2} + \Psi (q_t^2,({\sigma ^2} + q_t^2/\delta )/\sigma _s^2)/\delta }}{{\sigma _s^2}} \hfill,
\end{gathered}
\end{equation}
where the formula for $u_t$ is the result of the following proposition.

\begin{proposition}
\label{prop_u}
The optimal weighting parameter ${u_t}$ that combines the GENP $\tilde x$ and the previous iteration result in the GENP-AMP is given by
\begin{equation}
\label{weight}
{u_t} = \frac{{{\sigma ^2} + q_t^2/\delta }}{{\sigma _s^2}} .
\end{equation}
\end{proposition}

\begin{pf}
The optimal $u_t$ should minimize the MSE between the original sparse signal and the un-thresholded estimation $\hat{x}_0^t$ in (\ref{amp_x}), which can be obtained by minimizing ${(\frac{{{u_t}}}
{{1 + {u_t}}})^2}\sigma_s^2 + {(\frac{1}
{{1 + {u_t}}})^2}({\sigma ^2} + \frac{{{q_t^2}}}{\delta })$ over ${u_t}$.
\end{pf}

Replacing $u$ in Eq. (\ref{npi}) by Eq. (\ref{weight}), ${\text{npi}}({q^2},u;\delta ,\sigma ,{\sigma _s})$ can be simplified into
\begin{equation}
\label{simpnpi}
{\text{npi}}({q^2}) = \frac{{\sigma _s^2({\sigma ^2} + {q^2}/\delta )}}{{\sigma _s^2 + {\sigma ^2} + {q^2}/\delta }}.
\end{equation}

The fixed point condition of the state evolution is
\begin{equation}
\label{fixed}
q_*^2 = \Psi (q_*^2,\frac{{{\sigma ^2} + q_*^2/\delta }}{{\sigma _s^2}}) = \text{mse}(\text{npi}({q_*^2});p,\alpha ).
\end{equation}

If we treat $\xi^2={\text{npi}}({q_*^2})$ as an unknown variable, plugging (\ref{fixed}) into (\ref{simpnpi}) yields a fixed-point equation for $\xi^2$,
\begin{equation}
\label{fixed_pnt_eq}
{\xi ^2} = \frac{{\sigma _s^2({\sigma ^2} + \text{mse}({\xi ^2};p,\alpha )/\delta )}}{{\sigma _s^2 + {\sigma ^2} + \text{mse}({\xi ^2};p,\alpha )/\delta }}
\equiv F({\xi ^2},\alpha ).
\end{equation}

The following result shows that with an appropriate choice of $\alpha$, the fixed-point equation has a unique solution, from which we can predict the final MSE performance of the GENP-AMP algorithm.

\begin{proposition}
\label{fixed point}
Let ${\alpha _{\min }} = {\alpha _{\min }}(\delta ,{\gamma_s})$ be the unique non-negative solution of the equation
\begin{equation}
\label{fixed_pnt_alpha}
(1 + {\alpha ^2})\Phi ( - \alpha ) - \alpha \phi (\alpha ) = \frac{\delta }
{2}\frac{{{{(\gamma_s^2 + 1)}^2}}}
{{\gamma_s^4}},
\end{equation}
where $\phi (z)$ and $\Phi (z)$ are defined after Eq. (\ref{supmse}), and $\gamma_s^2$ is defined in Eq. (\ref{gamma}). Then for any $\alpha  > {\alpha _{\min }}(\delta ,{\gamma_s})$, the fixed-point equation ${\xi ^2} = F({\xi ^2},\alpha  )$ in (\ref{fixed_pnt_eq}) admits a unique solution ${\xi _*} = {\xi _*}(\alpha )$, and ${\lim _{t \to \infty }}{\xi _t} = {\xi _*}(\alpha )$.
\end{proposition}

\begin{pf}
This proof is an extension of Case $\chi  =  \pm $ in Appendix C of \citep{AMPoriginal}. It is easy to find that if $\gamma _s^2$ goes to $\infty $, the whole equation is exactly the one in \citep{Graphical}.

Since we want to have $F < {\xi ^2}$, following the same setup as the one in Case $\chi  =  \pm $ in Appendix C of \citep{AMPoriginal}, we need to consider the boundary point, which can be found by solving the boundary condition $\frac{{dF}}{{d{\xi ^2}}}{|_{{\xi ^2} = 0}} = 1$. This leads to $\frac
{\sigma_s^4 d(\Psi / \delta ) / d \xi^2}
{{{{({\sigma_s^2} + \sigma^2 + \Psi /\delta )}^2}}}{|_{{\xi ^2} = 0}} = 1$. If ${\xi ^2} \to 0$, we know that ${q^2}/\delta  = 0$, and the expression of $\frac{{d({q^2}/\delta )}}{{d{\xi ^2}}}$ can be obtained as in \citep{AMPoriginal}. Then the problem is transformed into
\begin{equation}
\label{diff_equ}
\frac{{d({q^2}/\delta )}}
{{d{\xi ^2}}}{|_{{\xi ^2} = 0}} = \frac{{{{(1 + \gamma _s^2)}^2}}}
{{\gamma _s^4}}.
\end{equation}
The numerator of Eq. (\ref{diff_equ}) becomes $\frac{{{{(1 + \gamma _s^2)}^2}}}
{{\gamma _s^4}}(1 - \frac{{\gamma _s^4}}
{{{{(1 + \gamma _s^2)}^2}}}\frac{2}
{\delta }[(1 + {\alpha ^2})\Phi ( - \alpha ) - \alpha \phi (\alpha )])$
 instead of $1 - \frac{2}
{\delta }[(1 + {\alpha ^2})\Phi ( - \alpha ) - \alpha \phi (\alpha )]$ as in the classical case in Eq. (6.6) of \citep{Graphical}. Comparing these two expressions, from Proposition 6.2 in \citep{Graphical}, we can reach the conclusion.
\end{pf}

If the threshold $\alpha$ and the distribution ${p_0}$ of ${X_0}$ are given, we can obtain the fixed point ${\xi _*}$ by solving Eq. (\ref{fixed_pnt_eq}). Therefore, the MSE performance of the GENP-AMP algorithm can be predicted.

Based on Prop. \ref{parameter}, $\lambda$ and $\tau_s $  can be determined if the necessary parameters are known. Conversely, if either $\lambda$ or $\tau_s$ is given, combining Eq. (\ref{fixed_pnt_alpha}) with Eq. (\ref{parameter_prediction}), we can get the corresponding $\alpha$ and ${\xi _*}$. Thus the other parameter  can be uniquely determined.

\section{Noise Sensitivity Analysis of GENP-AMP}
\label{noise_sensitivity}

The noise sensitivity phase transition is a curve in the $(\delta, \rho)$ plane \citep{AMP}, where $\rho=k/m$ and $\delta=m/n$, as defined in (\ref{kmn_def}). For many classical compressed sensing algorithms, the MSE is bounded below the phase transition curve, and unbounded above the curve. It is known that the optimal phase transition can be achieved by methods such as the AMP \citep{AMP}. ${\ell_1}$-based methods (such as the CVX package \citep{cvx}) can also have good phase transition performance. For large-scale problems, the OWLQN algorithm in \citep{OWLQN} has similar empirical phase transition boundary to ${\ell _1}$ methods, but its complexity is higher.

In this section, we show that there is no phase transition boundary for GENP-AMP, {\it i.e.}, its MSE is bounded in the entire plane, thanks to the GENP. We also prove that $\rho<1$ is no longer needed, which agrees with Lemma $1$ in \citep{Hui} for the elastic net-regularized LASSO.

First, for the GENP-LASSO problem in (\ref{minimization}), we define the MSE per entry when the empirical distribution of the signal converges to ${p_0}$:
\begin{equation}
{\text{MSE}}({\sigma ^2};\sigma _s^2,{p_0},\lambda ,{\tau _s}) = \mathop {\lim }\limits_{n \to \infty } \frac{1}{n}E\{ \left\| {\hat x(\lambda ,{\tau _s}) - {x_0}} \right\|_2^2\}  ,
\end{equation}
where the limit is taken along a converging sequence. Since the class ${\mathcal{F}_{\epsilon}}$ in (\ref{F_ep}) is scale-invariant, where $\varepsilon=k/n=\rho \delta$ according to (\ref{kmn_def}), the minimax risk of the GENP-LASSO can be written as
\begin{equation}
\label{risk}
\mathop {\inf }\limits_{\lambda ,\tau_s } \mathop {\sup }\limits_{{p_0} \in {F_{\rho \delta }}} {\text{MSE}}({\sigma ^2};\sigma_s^2,{p_0},\lambda ,\tau_s ) = {M^*}(\delta ,\rho ,\gamma_s^2){\sigma ^2},
\end{equation}
which indicates the sensitivity of the GENP-LASSO to the noise variance in the measurements, where $\gamma _s^2$ is defined in Eq. (\ref{gamma}), and the expression of noise sensitivity $M^*(\delta, \rho, \gamma _s^2)$ is given by the following proposition. We also give closed-form expressions of the tuning parameters $\lambda$ and ${\tau _s}$ that achieve the minimax risk bound.


Before presenting the proposition, we first define the formal mean square error (fMSE) and formal noise-plus interference level (fNPI), following Definitions $3.1{-}3.4$ in \citep{AMP}. fMSE is defined as the MSE of an observable in a large system framework ${\text{LSF}}(\delta ,\rho ,\sigma ,{\gamma _s},p)$, where ${\text{LSF}}(\delta ,\rho ,\sigma ,{\gamma _s},p)$ denotes a sequence of problem instances ${(y;A,x)_{m,n}}$ as per Eq. (\ref{x_model}) indexed by the problem sizes, and $m$ and $n$ grow proportionally such that $m/n=\delta $. fNPI is expressed as
\begin{equation}\nonumber
\begin{split}
{\text{fNPI}} &= {(\frac{{{u^*}}}
{{1 + {u^*}}})^2}\sigma_s^2 + {(\frac{1}
{{1 + {u^*}}})^2}({\sigma ^2} + {\text{fMSE}}/\delta ), \\
{u^*} &= \frac{{{\sigma ^2} + {\text{fMSE}}/\delta }}
{{\sigma_s^2}}.
\end{split}
\end{equation}
Its minimax value is ${\text{NP}}{{\text{I}}^*}(\delta ,\rho ,{\gamma_s^2}) \equiv {\text{ }}\frac{{\gamma_s^2{\sigma ^2}(1 + {M^*}(\delta ,\rho ,{\gamma_s^2})/\delta )}}
{{\gamma_s^2 + 1 + {M^*}(\delta ,\rho ,{\gamma_s^2})/\delta }}$ by replacing ${\text{fMSE}}$ in the equation above with its minimax risk ${M^*}(\delta ,\rho ,\gamma _s^2)$.

\begin{proposition}
\label{phase transition}
(1) For any point in the surface, i.e., $\rho  \leqslant 1/\delta $ (since $\delta \rho  = \varepsilon  \leqslant 1$), the minimax risk of GENP-LASSO is bounded, and $M^*(\delta, \rho, \gamma _s^2)$ is given by
\begin{equation}
\label{phase_tx_mstar}
\small{
{M^*}(\delta ,\rho ,\gamma _s^2) = \frac{{ - G(\delta ,\rho ,\gamma _s^2) + \sqrt {G{{(\delta ,\rho ,\gamma _s^2)}^2} + 4\delta \gamma _s^2{M^ \pm }(\delta \rho )} }}{2},
}
\end{equation}
where $G(\delta ,\rho ,{\gamma_s^2}) = \delta \gamma_s^2 + \delta  - \gamma_s^2{M^ \pm }(\delta \rho )$.

(2)For $c > 0$, define
\begin{equation}\nonumber
{h^*}(\delta ,\rho ,{\gamma_s^2};c){\text{ }} \equiv {\text{ }}{h^ \pm }(\delta \rho ,c){\text{ }} \cdot {\text{ }}\sqrt {{\text{NP}}{{\text{I}}^*}}.
\end{equation}
Then similar to Eq. (\ref{least_favor_p}), the distribution $p \in {\mathcal{F}_{\delta \rho }}$ with a fraction $(1 - \delta \rho)$ of its mass at zero and the remaining mass equally at $ \pm {h^*}(\delta ,\rho ,{\gamma_s^2};c)$ is $c$-nearly-least-favorable, {\it i.e.}, the formal noise sensitivity of $\hat x(\lambda ,{\tau _s})$ is
\beq
\small{
\frac{{ - G(\delta ,\rho ,\gamma _s^2;c) + \sqrt {G{{(\delta ,\rho ,\gamma _s^2)}^2} + 4(1 - c)\delta \gamma _s^2{M^ \pm }(\delta \rho )} }}{2},
}
\eeq
where $G(\delta ,\rho ,{\gamma_s};c) = \delta \gamma_s^2 + \delta - (1 - c){M^ \pm }(\delta \rho )\gamma_s^2$.

(3) The formal minimax parameters are given by
\begin{equation}
\begin{split}
  \lambda (\upsilon ;\delta ,\rho ,\sigma ,{\sigma_s}) &\equiv (1 + {u^*}) \cdot {\alpha ^ \pm }(\delta \rho ) \cdot \sqrt {{\text{fNPI}}({\alpha ^ \pm };\delta ,\rho ,\sigma ,{\sigma_s},\upsilon )} \\
& \times (1 - \frac{1} {{1 + {u^*}}}{\text{EqDR}}(\upsilon ;{\alpha ^ \pm }(\delta \rho ))/\delta ), \hfill \\
  \tau_s (\upsilon ;\delta ,\rho ,\sigma ,{\sigma_s}) &\equiv {u^*}(1 - \frac{1}
{{1 + {u^*}}}{\text{EqDR}}(\upsilon ;{\alpha ^ \pm }(\delta \rho ))/\delta ),
\end{split}
\end{equation}
where $\text{EqDR}$ is the equilibrium detection rate, {\it i.e.}, the asymptotic fraction of coordinates that are estimated to be nonzero, {\it i.e.}, ${\text{EqDR}} = P\{ \eta ({x_\infty };{\theta _\infty }) \ne 0\} $, as in Eq. (4.5) in \citep{AMP}.
\end{proposition}

\begin{pf}
The proof is given in Appendix \ref{apx_phase}.
\end{pf}

To show that the noise sensitivity analysis presented here is indeed a generalized result, we next discuss three special cases and show that the result here degrades to the existing known conclusions. First, let $\gamma _s^2 = \infty $. In this case, Eq. (\ref{phase_tx_mstar}) degrades to the formulae of the bounded MSE below the phase transition boundary of AMP, {\it i.e.}, Eq. (4.8) in \citep{AMP} . The phase transition boundary only exists in this extreme case for GENP-AMP. Second, if $\gamma _s^2 = 0$, {\it i.e.}, $\tilde{x}=x$, we do not need to run the AMP; hence the MSE is 0, which coincides with Eq. (\ref{phase_tx_mstar}) when $\gamma _s^2 = 0$. Last, if $\delta  = 0$, which means there is no compressed measurement, solving the minimization problem in Eq. (\ref{minimization}) is equivalent to scalar denoising, and the minimax MSE is ${M^ \pm }(\rho \delta )\sigma _s^2$, which also agrees with the denoising of scalars introduced in Sec. \ref{sec_bkgd}.

When there is no initial estimation $\tilde{x}$, the formal MSE noise sensitivity above the phase transition is infinite. However, this is no longer the case in the presence of the GENP, as we can at least assign $\tau_s$ to $\infty $ while keeping $\lambda$ to be finite, and the formal MSE noise sensitivity is thus bounded by $\gamma_s^2$. We can do even better by exploiting the measurement and the sparsity of the original signal, as shown below.

 It is easy to verify that $\partial {M^*}(\delta ,\rho ,\gamma _s^2) / \partial \gamma _s^2$ is positive, so ${M^*}(\delta ,\rho ,\gamma _s^2)$ is a monotonically increasing function of $\gamma _s^2$. Since GENP-AMP reduces to AMP when $\gamma_s^2=\infty $, this means that the minimax bound of GENP-LASSO  is no greater than that of LASSO, {\it i.e.},
\begin{equation}
\label{TheorEvid1}
{M^*}(\delta ,\rho ,\gamma_s^2) \leqslant {M^b}(\delta ,\rho ),
\end{equation}
where ${M^b}(\delta ,\rho ) = \frac{{{M^ \pm }(\delta \rho )}}{{1 - {M^ \pm }(\delta \rho )/\delta }}$ is the bound of LASSO minimax risk.

Besides, we can also verify that for a fixed sparsity, i.e., $\varepsilon  = \delta \rho $ is a constant, $\partial {M^*}(\delta ,\rho ,\gamma _s^2)/\partial \delta $ is non-positive (only equal to 0 when $\delta  = 0$), and ${M^*}(\delta ,\rho ,\gamma _s^2)$ is a monotonically decreasing function of $\delta $. Since GENP-AMP reduces to denoising via soft-thresholding described in Sec. \ref{sec_bkgd} when $\delta  = 0$, we conclude that the minimax bound of GENP-LASSO is no greater than that of scalar denoising,
\begin{equation}
\label{TheorEvid2}
{M^*}(\delta ,\rho ,\gamma _s^2) \leqslant {M^ \pm }(\delta \rho )\gamma _s^2.
\end{equation}

In fact, Eq. (\ref{TheorEvid1}) and  (\ref{TheorEvid2}) have proved that GENP-AMP outperforms AMP and the scalar denoising via soft-thresholding. More importantly,  Eq. (\ref{TheorEvid1}) measures the benefit brought by the generalized elastic net prior while Eq. (\ref{TheorEvid2}) measures the benefit brought by the linear CS measurements.

We can find more properties of this minimax risk bound. For a fixed $\delta$, the only function of $\rho $ is ${M^ \pm }(\delta \rho )$. From \citep{AMP}, we know that ${M^ \pm }(\delta \rho )$ is monotonically increasing with respect to $\rho $, and ${M^ \pm }(0) \to 0,{\text{ }}{M^ \pm }(1) \to 1$. Besides, we can find that ${M^*}(\delta ,\rho ,\gamma _s^2)$ is monotonically increasing with respect to ${M^ \pm }(\delta \rho )$. The maximum value of ${M^ \pm }(\delta \rho )$  is 1. The maximum value of  ${M^*}(\delta ,\rho ,\gamma _s^2)$ is thus
\begin{equation}
\label{maximum minimax risk}
\begin{split}
& \mathop {\max } \limits_{{M^ \pm }(\delta \rho )} {M^*}(\delta ,\rho ,\gamma _s^2) \\
& = \frac{{\sqrt {{{(\delta \gamma _s^2 - \gamma _s^2 + \delta )}^2} + 4\delta \gamma _s^2}  - (\delta \gamma _s^2 - \gamma _s^2 + \delta )}}
{2},
\end{split}
\end{equation}
where the maximum is achieved at $\rho  = 1/\delta $.

\section{Parameterless GENP-AMP}
\label{Parameterless}


In the GENP-AMP proposed above, two parameters need to be known in advance: (1) the sparsity of the signal, $\varepsilon  = k/n$, in order to select the appropriate thresholding parameter in soft thresholding function in Sec. \ref{sec_bkgd}; (2) the variance of the prior $\tilde x$, $\sigma _s^2$, in order to determine the weighting parameter ${u_t}$ as in Prop. \ref{prop_u}. This makes the algorithm impractical.

The original AMP also needs to know the sparsity. However, recently two types of parameterless AMP algorithms have been developed in \citep{Paraless} and \citep{EMAMP, Vila132}. In \citep{Paraless}, Stein's unbiased risk estimate (SURE) framework is used to automatically determine the optimal thresholding parameter in AMP using the gradient descent method. The methods in \citep{EMAMP,Vila132} are both based on the GAMP \citep{GAMP}, and try to approximate the MMSE result by learning the prior distribution of the sparse signal through Expectation Maximization (EM) method.

In this part, we follow the approach in \citep{Paraless} due to its theoretical guarantee, since the complete analysis of the EM algorithm used in \citep{EMAMP,Vila132} is still not available. However, the method in \citep{Paraless} cannot be applied in this paper directly since it does not consider the GENP. In the following proposition, using the SURE theory, we develop a practical parameterless version of the GENP-AMP (P-GENP-AMP) that can simultaneously select the thresholding parameter and estimate the variance of the GENP.

\begin{proposition}
\label{gamma_appro}
 The variance of the GENP $\tilde x$ can be approximated by
 \begin{equation}
 \label{var_true}
\sigma _s^2 \approx \frac{{\left\| {\tilde x - {x_{{\text{AMP}}}}} \right\|_2^2 - \mathop {\lim }\limits_{t \to \infty } \hat r({\theta ^t})}}{n},
\end{equation}
where ${x_{{\text{AMP}}}}$ is the sparse signal estimated by the AMP with the same setup (fixed $A$, $\delta$, and $\rho $), ${\mathop {\lim }\limits_{t \to \infty } \hat r({\theta ^t})}/{n}$ is the MSE of AMP predicted by the SURE method in \citep{Paraless}, and
\begin{equation}
\label{SUREappro}
\frac{{\hat r({\theta _t})}}{n} = \frac{1}{n}\left\| {\eta (\hat x_0^t;{\theta _t}) - \hat x_0^t} \right\|_2^2 + \sigma _t^2 + \frac{1}{n}\sigma _t^2[{1^T}(\eta '(\hat x_0^t;{\theta _t}) - 1)]
\end{equation}
 is Eq. (13) in \citep{Paraless}, in which $\sigma _t^2$ is the noise-plus interference level in the $t$th iteration of the standard AMP.
\end{proposition}

\begin{pf}
The proof is given in Appendix \ref{gamma_proof}.
\end{pf}

In fact, thanks to the state evolution analysis, the choice of ${x_{{\text{AMP}}}}$ can be quite flexible. Another good choice is $\hat x_0^*$, the un-thresholded estimator in the last iteration of AMP, whose variance is $\sigma _*^2$, mentioned in Eq. (\ref{amp_01}). Then, $\sigma _s^2$ can also be approximated by
\begin{equation}
\sigma _s^2 \approx \frac{{\left\| {\tilde x - \hat x_0^*} \right\|_2^2 - \sigma _*^2}}{n}.
\end{equation}

Note that as shown in Prop. \ref{gamma_appro} and its proof in Appendix \ref{gamma_proof}, the approximation of $\sigma _s^2$ relies on the approximation of the standard AMP. Therefore, above the phase transition boundary of AMP, the AMP approximation is unstable since the MSE is unbounded, making the approximation $\mathop {\lim }\limits_{t \to \infty } \hat{r} (\theta ^t)/n$ unbounded. A tiny mismatch between $\mathop {\lim }\limits_{t \to \infty } \hat{r} (\theta ^t)/n$ and MSE of AMP will cause large error when estimating $\sigma _s^2$. On the other hand, below the phase transition boundary, the MSE of AMP is bounded. The approximation is very stable.

Once $\sigma _s^2$ is estimated, the remaining problem is to determine the thresholding parameter in Eq. (\ref{amp_x}). Since the iteration formulae and the state evolutions of GENP-AMP are similar to those of AMP, we only need to replace the explicit expressions of $\sigma _t^2$ in Eq. (\ref{SUREappro}) with ${\text{npi(}}q_t^2)$ in Eq. (\ref{simpnpi}). The subsequent steps are exactly the same as those in \citep{Paraless}, {\it i.e.}, determining the thresholding parameter ${\theta _t}$ using gradient descent, and updating the estimator and the residual according to Eq. (\ref{amp_x}) and (\ref{amp_r}).

\begin{figure}[tb]
\begin{center}
\begin{tabular}{c}
  \includegraphics[width=2.7in]{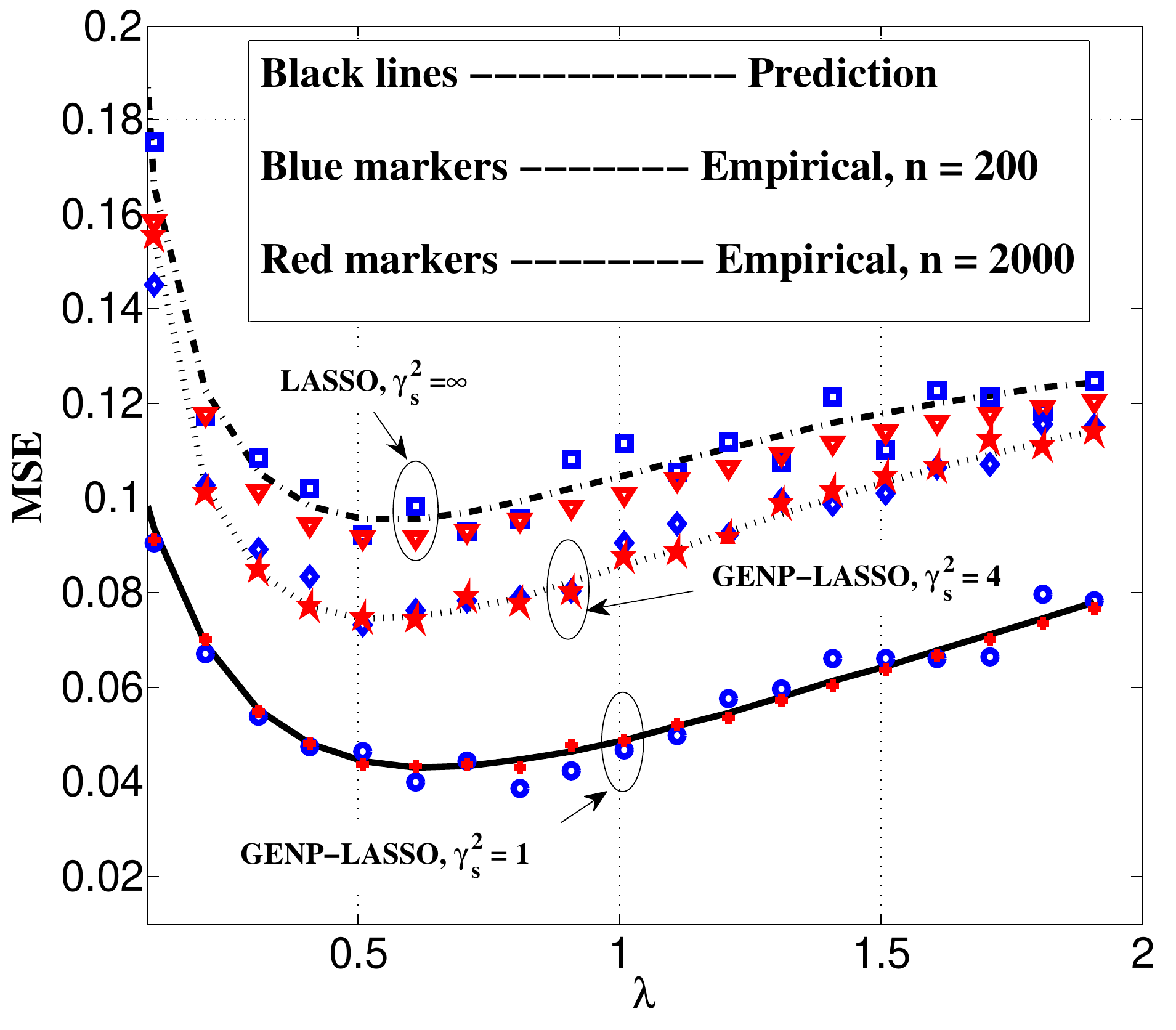}
\end{tabular}
\end{center}
\vskip -5pt
\caption{\label{prediction} The predicted and actual MSEs of LASSO and GENP-LASSO with different regularization parameter $\lambda$. The sample rate is $\delta = 0.64$.}
\vskip -10pt
\end{figure}


\section{Numerical Experiments}
\label{sec_exp}

In this section, we present simulation results with 1-D data and two different imaging applications to demonstrate the performances of the proposed GENP-LASSO and GENP-AMP. Comparisons with some other methods are also included.

\subsection{Performance of GENP-LASSO}

We first compare the predicted and empirical MSEs of GENP-LASSO and LASSO. Note that GENP-LASSO reduces to LASSO when ${\gamma_s^2} = \infty$. We generate the signal vector ${x_0}$ by randomly choosing each entry from $\{  + 1,0, - 1\}$ with probabilities $P({x_{0,i}} =  + 1) = P({x_{0,i}} =  - 1) = 0.064$. The entries of the measurement matrix $A$ are drawn from the i.i.d. Gaussian distribution $\mathcal{N}(0,1/m)$. The sampling noise $w$ are drawn from $\mathcal{N}(0,0.2)$, and the noise $e$ of the GENP $\tilde x$ are drawn from $\mathcal{N}(0,0.2\gamma_s^2)$. The simulation setup is the same as that in \citep{Graphical}, except for the GENP.

As shown in Sec. \ref{sec_siamp}, the MSE of GENP-LASSO is controlled by two regularization parameters $\lambda$ and $\tau_s $, but they are connected by the hidden parameter $u$. If one of them is given, using Prop. \ref{parameter}, Prop. \ref{prop_u}, and Prop. \ref{fixed point}, the other parameters can be uniquely determined.

Fig. \ref{prediction} shows the predicted and the empirical MSEs of LASSO and GENP-LASSO with different $\lambda$. Three $\gamma_s^2$ are tested, each with two different values of $n$. In this example, the predicted MSEs of GENP-LASSO are given by the state evolution of GENP-AMP. The empirical results of LASSO and GENP-LASSO for $n=200$ are obtained by the Matlab-based CVX package \citep{cvx}. The empirical results of LASSO for $n=2000$ are obtained by the OWLQN algorithm \citep{OWLQN}, which is written in C++. The empirical results of GENP-LASSO for $n=2000$ are obtained by modifying the OWLQN to incorporate the GENP, as described in Sec. \ref{sec_inf}. We denote this as GENP-OWLQN.

It can be seen from Fig. \ref{prediction} that the predicted MSE is quite accurate in both LASSO and GENP-LASSO. The result of LASSO (with ${\gamma_s^2} = \infty $) is the same as Fig. 9 in \citep{Graphical}. When ${\gamma_s^2} = 4$ or ${\gamma_s^2} = 1$, the minimal MSE of GENP-LASSO can be reduced by about $20 \%$ and $50 \%$, respectively, compared to the standard LASSO without any prior.

\begin{table*}[!tb]
\centering
\vskip 10pt
\scalebox{0.6}{
\begin{tabular}{c|c|c|c|c|c|c|c|c|c|c|c|c|c}
\hline
$\delta $ & $\rho $ & ${h^*}$ & ${\lambda ^*}$ & ${\tau ^*}$ & fMSE & eMSE &  eMSE & fMSE & eMSE & eMSE & fMSE & eMSE & eMSE \\
&&&&& (GENP & (GENP- & (GENP & (AMP) & (OWLQN) & (AMP) & (DN) & (DN) & (LMMSE)\\
&&&&& -AMP) & OWLQN) & -AMP) &  &  &  &  & & \\
\hline
0.100 & 0.095 & 2.828 & 2.585 & 0.995 & {0.033}  & 0.032 & 0.033  & 0.136 & 0.119 & 0.128 & 0.058 & 0.062 & 0.071 \\

0.100 & 0.142 & 2.807 & 2.359 & 0.993 & {0.047} & 0.044 & 0.048 & 0.380 & 0.394 & 0.430 & 0.079 & 0.081 & 0.098\\

0.100 & 0.170 & 2.801 & 2.256 & 0.992 & {0.055} & 0.057 & 0.056 & 1.045 & 1.199 & 1.089 & 0.090 & 0.093 & 0.111 \\

0.100 & 0.180 & 2.799 & 2.223 & 0.992 & {0.058} & 0.058 & 0.058 & 2.063 & 1.958 & 3.159 & 0.094 & 0.103 & 0.116 \\

0.100 & 1.900 & 2.656 & 0.919 & 0.951 & {0.405} & 0.405 & 0.406 & UB & UB & UB & 0.486 & 0.479 & 0.525\\

\hline
0.250 & 0.134 & 2.581 & 2.025 & 0.995 & {0.086} & 0.091 & 0.088 & 0.374 & 0.369 & 0.366 & 0.150 & 0.151 & 0.167\\

0.250 & 0.201 & 2.547 & 1.796 & 0.994 & {0.120} & 0.121 & 0.123 & 1.028 & 1.213 & 1.137 & 0.201 & 0.203 & 0.213\\

0.250 & 0.241 & 2.533 & 1.694 & 0.993 & {0.139} & 0.137 & 0.139 & 2.830 & 2.708 & 2.910 & 0.228 & 0.226 & 0.243 \\

0.250 & 0.254 & 2.529 & 1.663 & 0.992 & {0.145} & 0.145 & 0.148 & 5.576 & 6.665 & 5.680 & 0.236 & 0.236 & 0.251 \\

0.250 & 1.900 & 2.276 & 0.511 & 0.973 & {0.619} & 0.625 & 0.626 & UB & UB & UB & 0.797 & 0.790 & 0.592 \\
 \hline
0.500 & 0.193 & 2.362 & 1.512 & 0.995 & {0.182} & 0.184 & 0.184 & 0.853 & 0.845 & 0.856 & 0.315 & 0.316 & 0.289\\

0.500 & 0.289 & 2.314 & 1.279 & 0.992 & {0.245} & 0.245 & 0.245 & 2.329 & 2.343 & 2.412 & 0.410 & 0.415 & 0.345 \\

0.500 & 0.347 & 2.291 & 1.172 & 0.993 & {0.280} & 0.275 & 0.280 & 6.365 & 7.232 & 6.312 & 0.459 & 0.465 & 0.367 \\

0.500 & 0.366 & 2.285 & 1.140 & 0.993 & {0.291} & 0.296 & 0.290 & 12.427 & 15.665 & 12.165 & 0.475 & 0.476 & 0.386\\

0.500 & 1.900 & 1.253 & 0.047 & 0.986 & {0.689} & 0.689 & 0.696 & UB & UB & UB & 0.978 & 0.972 & 0.458 \\
\hline
\end{tabular}
}
\caption{\label{validation} Empirical and predicted MSEs of different methods for different points in the sampling space.}
\vskip -15pt
\end{table*}

\subsection{Comparison of AMP, GENP-AMP, Denoising and Least Squares}

We now compared the performances of AMP, GENP-AMP, the LMMSE solution for Eq. (\ref{eq_ls}), and scalar denoising via soft thresholding of the initial estimation when they are operated at different points of the sampling plane, including points below and above the phase transition boundary of the standard AMP. We will compare the predicted and empirical MSEs of GENP-AMP and AMP using the nearly-least-favorable signal generated by Eq. (\ref{least_favor_p}). We also use OWLQN and GENP-OWLQN to find the LASSO solution $\hat x(\lambda )$ and the GENP-LASSO solution $\hat x(\lambda ,{\tau _s})$ for Eq. (\ref{minimization}), but OWLQN-based methods could not predict the MSE, and the regularized parameters need to be chosen manually. The number of iterations of GENP-AMP and AMP for empirical results is fixed as $60$.

We first generate in each case $20$ random realizations of size $n =2000$, with parameters , $\gamma _s^2 = 1$, ${\sigma ^2} = 1$, $\delta  \in \{ 0.10,0.25,0.50\} $, $\rho  \in $ $\{ \frac{1}{2}\rho (\delta ),\frac{3}{4}\rho (\delta ),\frac{9}{{10}}\rho (\delta ), \\
\frac{{19}}{{20}}\rho (\delta ),1.9\} $, where $\rho (\delta )$ represents the phase transition boundary of the standard AMP. The results are summarized in Table \ref{validation}, where eMSE and fMSE denote the empirical MSE and predicted formal MSE respectively. DN denotes the denoising method, and UB represents unbounded MSE. More results with different $\gamma _s^2$ are shown in Table \ref{gammas}.

\begin{table*}[!tb]
\centering
\vskip 10pt
\scalebox{0.75}{
\begin{tabular}{c|c|c|c|c|c|c|c|c|c|c|c}
\hline
$\gamma _s^2$  & $\delta $ & $\rho $ & ${h^*}$ & ${\lambda ^*}$ & ${\tau ^*}$ & fMSE & eMSE &  eMSE  & fMSE & eMSE & eMSE \\
&&&&&& (GENP & (GENP- & (GENP & (DN) & (DN) & (LMMSE)\\
&&&&&& -AMP) & OWLQN) & -AMP) &  &  & \\
\hline
\multirow{5}{*}{2} & 0.100 & 0.095 & 3.465 & 2.107 & 0.497 & 0.049  & 0.047 & 0.047  & 0.115 & 0.105 & 0.108 \\

& 0.100 & 0.142 & 3.511 & 1.882 & 0.495 & {0.073} & 0.077 & 0.077 & 0.157 & 0.134 & 0.145\\

& 0.100 & 0.170 & 3.539 & 1.779 & 0.494 & {0.087} & 0.086 & 0.086 & 0.181 & 0.161 & 0.173 \\

& 0.100 & 0.180 & 3.549 & 1.747 & 0.494 & {0.093} & 0.093 & 0.094 & 0.189 & 0.165 & 0.189 \\

& 0.100 & 1.900 & 3.717 & 0.625 & 0.452 & {0.794} & 0.807 & 0.808 & 0.971 & 0.870 & 1.030\\
\hline
\multirow{5}{*}{4} & 0.100 & 0.095 & 4.086 & 1.785 & 0.248 & 0.068  & 0.070 & 0.071  & 0.231 & 0.148 & 0.140 \\

& 0.100 & 0.142 & 4.271 & 1.543 & 0.246 & {0.108} & 0.114 & 0.115 & 0.315 & 0.205 & 0.234\\

& 0.100 & 0.170 & 4.377 & 1.433 & 0.245 & {0.133} & 0.128 & 0.129 & 0.361 & 0.242 & 0.289 \\

& 0.100 & 0.180 & 4.413 & 1.398 & 0.245 & {0.142} & 0.148 & 0.148 & 0.377 & 0.250 & 0.291 \\

& 0.100 & 1.900 & 5.224 & 0.399 & 0.203 & {1.566} & 1.566 & 1.567 & 1.942 & 1.459 & 2.046\\
 \hline
\end{tabular}
}
\caption{\label{gammas} Empirical and predicted MSEs of different methods with different $\gamma _s^2$.}
\vskip -15pt
\end{table*}

Some observations can be drawn from Tables \ref{validation} and \ref{gammas}. First, the MSE of GENP-AMP is much lower than those of AMP and denoising. Secondly, the fMSE and eMSE of GENP-AMP match very well, even when the number of measurements is smaller than the sparsity. For example, for $\rho= 1.9$, the fMSE of GENP-AMP is still very close to eMSE. For AMP, this $\rho$ is much higher than its phase transition boundary. Its MSE is thus unbounded. Thirdly, since the denoising method is equivalent to GENP-AMP with $\delta  = 0$, the performance difference between GENP-AMP and denoising shows the contribution of the CS measurements. Moreover, the LMMSE solution is comparable to DN solution. The exceptions happen when $\delta  = 0.25,{\text{ }}\rho  = 1.9$ and $\delta  = 0.5,{\text{ }}\rho  = 1.9$. This can be expected since LMMSE can be interpreted as assuming the target signal $x$ follows Gaussian distribution. When $\varepsilon  = \delta \rho $ is sufficiently large, the distribution of $x$ is close to Gaussian distribution, according to central limit theorem. In this case, the LMMSE result is near-optimal. Especially, when $\delta  = 0.5,{\text{ }}\rho  = 1.9$, i.e., $\varepsilon  = 0.95$, almost all entries of $x$ are nonzero, LMMSE outperforms other methods. However, in all other cases, LMMSE is worse than our proposed algorithm.

Finally, although the empirical MSE of GENP-OWLQN is very similar to that of GENP-AMP, GENP-OWLQN is much slower, since it needs to calculate the gradients in each iteration. For example, on a computer with Intel Core i7 3.07GHz CPU and 6.00 GB memory, our Matlab implementation of GENP-AMP is about $10$ times faster than the C++ implementation of GENP-OWLQN.

\begin{figure*}[t]
\begin{center}
    \begin{tabular}{c@{\hspace{-0.1mm}}c@{\hspace{-0.1mm}}c@{\hspace{-0.1mm}}c}
	\includegraphics[width=2.3in]{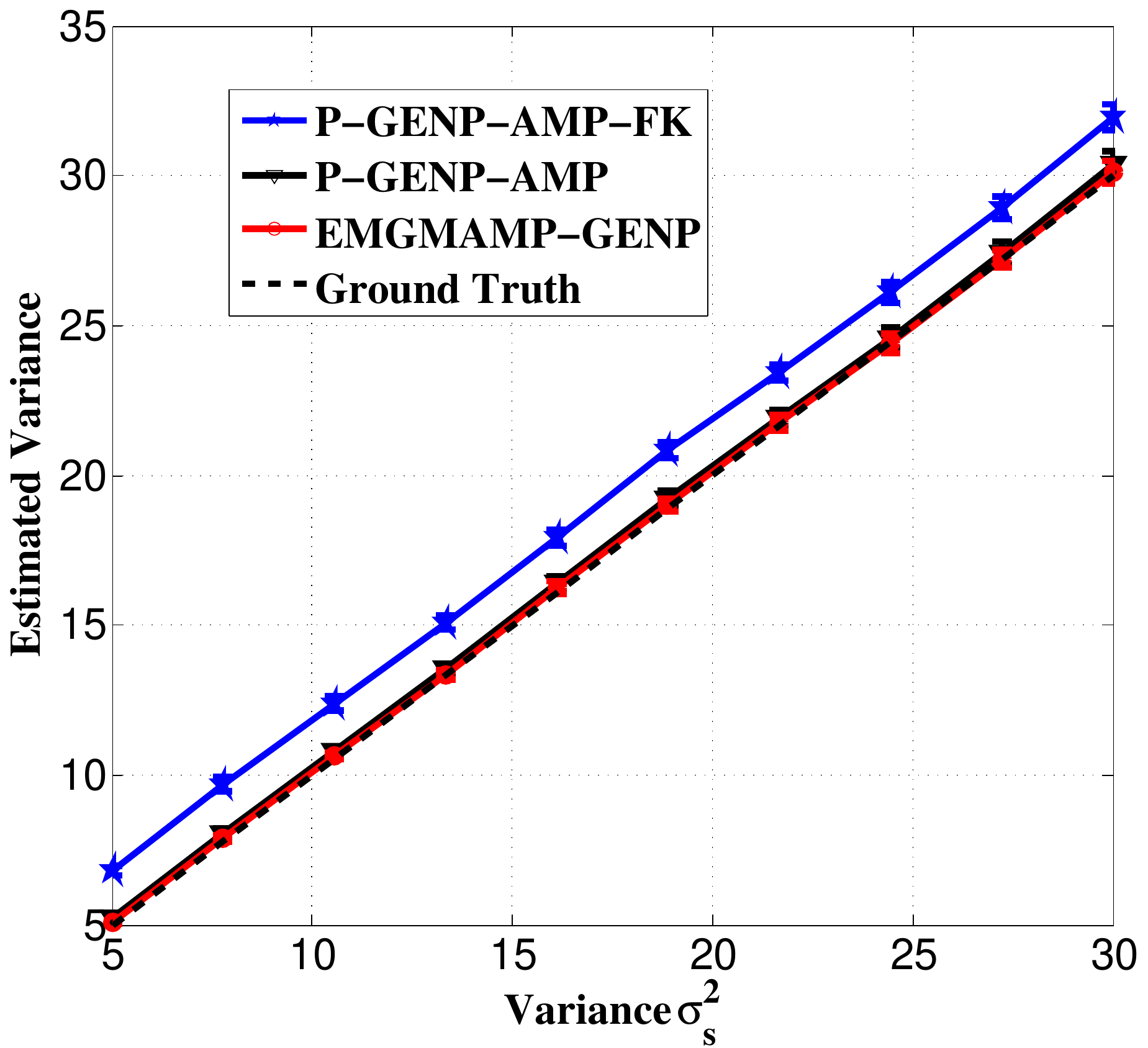} &
	\includegraphics[width=2.3in]{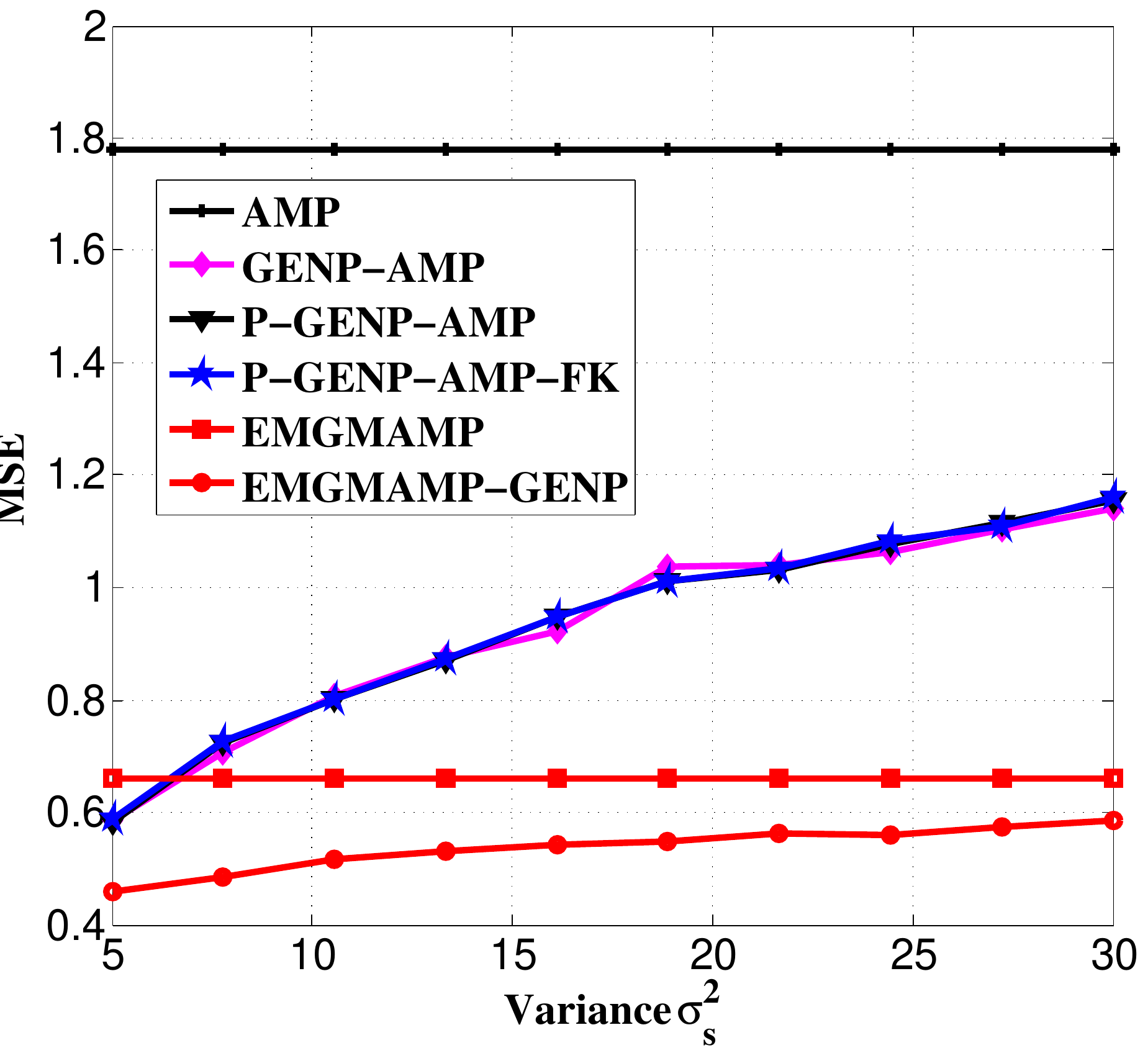} \\
	\includegraphics[width=2.3in]{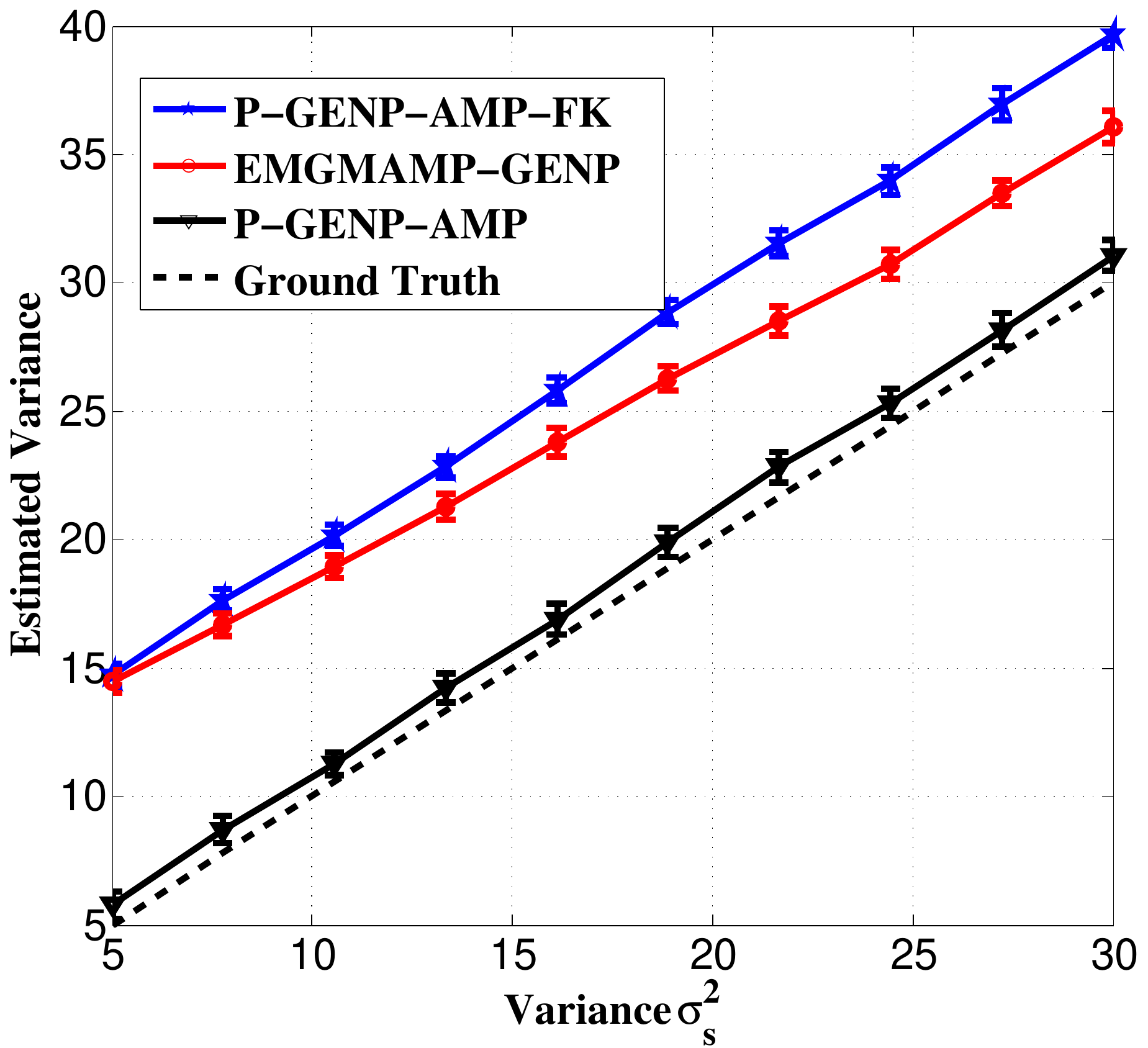} &
	\includegraphics[width=2.3in]{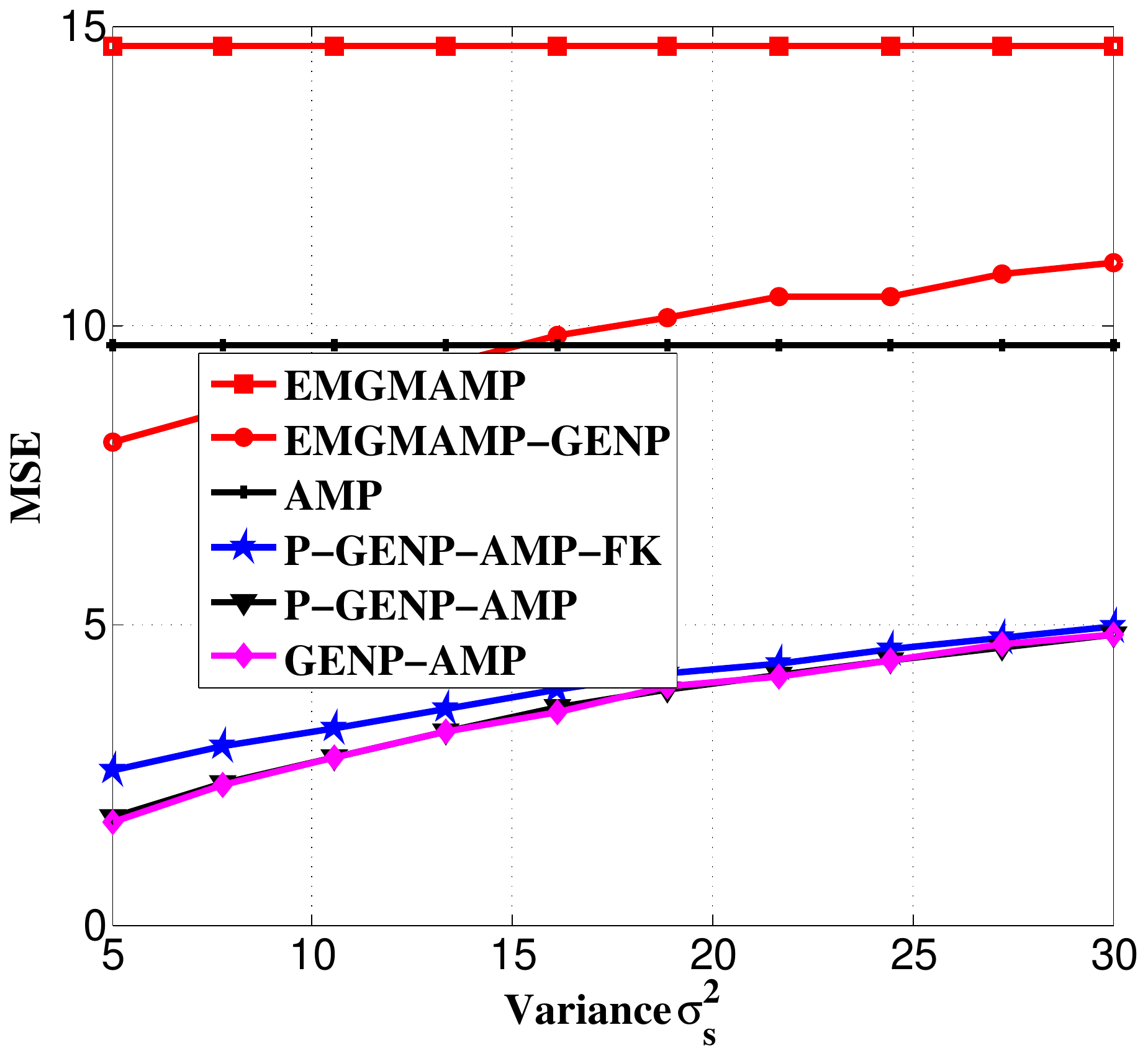} \\
    \end{tabular}
\vskip -5pt
\caption{\label{paraless_perf}Performances of parameterless algorithms with $\delta  = 0.5$ and $\varepsilon  = 0.2$. First row (from left to right): (a) Estimated $\sigma_s^2$ with SNR=20 dB. The confidence level of the error bar is 0.95. (b) MSEs with SNR=20 dB.  Second row: (c) Estimated $\sigma_s^2$ with SNR=5 dB. The confidence level of the error bar is 0.95. (d) MSEs with SNR=5 dB. }
\end{center}
\vskip -15pt
\end{figure*}

\subsection{Performance of the Parameterless GENP-AMP}
\label{PGENPAMPsimul}

In the previous two simulations, the sparsity $\varepsilon$ and the variance $\sigma _s^2$ of the prior $\tilde x$ are assumed to be known. In this subsection, we show the performance of the parameterless GENP-AMP (P-GENP-AMP), which can estimate $\sigma _s^2$. A similar setup to the previous experiments is used, except for the following. The non-zero coefficients of the sparse signal $x$ follow i.i.d. $\mathcal{N}(0,100)$.  The sampling noise $w$ are drawn from $\mathcal{N}(0,{\sigma ^2})$ where the variance ${\sigma ^2}$ is set according to signal-to-noise ratio (SNR) defined as ${\text{SNR}} = 10{\log _{10}}(\frac{1}{m}\left\| {Ax} \right\|_2^2/{\sigma ^2})$, and the noise $e$ of the GENP $\tilde x$ are drawn from $\mathcal{N}(0,\sigma _s^2)$.  The number of Monte-Carlo simulations is 100.

For comparison purpose, we also estimate $\sigma_s^2$ using the following method
\begin{equation}
\label{var_fake}
\sigma _s^2 \approx \frac{1}{n}\left\| {\tilde x - {x_{{\text{AMP}}}}} \right\|_2^2,
\end{equation}
{\it i.e.}, we first reconstruct the sparse signal using standard CS reconstruction methods such as AMP, and then use the reconstructed signal and $\tilde{x}$ to estimate $\sigma_s^2$. And we name such kind of algorithm as Parameterless GENP-AMP with faked variance (P-GENP-AMP-FK). In fact, the only difference between Eq. (\ref{var_true}) and Eq. (\ref{var_fake}) is the term ${{\mathop {\lim }\limits_{t \to \infty } \hat r({\theta ^t})}}/{n}$, the estimated MSE by the SURE framework proposed in \citep{Paraless}.

We also compare with the method in \citep{EMAMP}, denoted as EMGMAMP, using its source code from \citep{gamptoolbox}. We modify its source code to incorporate the GENP, and treat the variance of GENP as an additional hidden parameter, which can also be updated by the Expectation-Maximization algorithm in \citep{EMAMP}. This algorithm is denoted as EMGMAMP-GENP in the following figures. The updating rule follows
\begin{equation}
\sigma _s^2(t) = \frac{1}{n}\sum\limits_{i = 1}^n {[{{({{\tilde x}_i} - {{\hat x}_i}(t))}^2} + \mu _i^x{{(t)}^2}]},
\end{equation}
where ${{{\hat x}_i}(t)}$ and ${\mu _i^x(t)}$ is the approximate MMSE result, and its standard deviation in the $t$-th iteration, respectively.

\begin{table*}[t]
\centering{
\vskip 10pt
\scalebox{0.85}{
\begin{tabular}{c*{9}{|c}}
\hline
 ${\sigma ^2},\sigma _s^2$ & $\delta$  & Alg1 & Alg2 & Alg3 & Alg4 & Alg5 & Alg6 & Alg7 & Alg8  \\
\hline
 {\multirow{2}{*}{$1{e3},1{e3}$}} & 1/5 & 24.73 & \textbf{26.34} & \textbf{26.40} & 24.48 & 23.24 & 17.79 & 24.95 & 17.83 \\
  & 1/2 & 26.15 & \textbf{26.87} & \textbf{26.86} & 20.58 & 20.04 & 18.14 & 24.95 & 26.49 \\
\cline{1-10}
{\multirow{2}{*}{$1{e3},2.5{e3}$}} & 1/5  & 24.73 & \textbf{25.91} & \textbf{25.89} & 24.48 & 23.97 & 14.00 & 24.15 & 6.73 \\
  & 1/2 & 26.15 & \textbf{26.57} & \textbf{26.61} & 20.58 & 19.75 & 14.70 & 24.15 & 26.03 \\
\hline
\end{tabular}
}
}
\vskip 5pt
\caption{\label{Upsampled} PSNRs of different methods for the reconstruction of "Lena". For $\sigma _s^2 = 1{e3}$, the PSNR of the corrupted upsampled version are all 18.13 dB, whereas when $\sigma _s^2 = 2.5{e3}$, the PSNR is 14.00 dB.}
\vskip -10pt
\end{table*}

In the first experiment, we consider a high SNR of 20 dB. From Fig. \ref{paraless_perf}(a), we can see that P-GENP-AMP, and P-GENP-AMP-FK can both provide good approximations of the variance $\sigma _s^2$ while the gap between the ones estimated by P-GENP-AMP and GENP-AMP is exactly the MSE of AMP shown in Fig. \ref{paraless_perf} (b). It can also be seen from Fig. \ref{paraless_perf} (b) that all GENP-based algorithms achieve better performances. EMGMAMP-GENP outperforms the others, since it can learn the prior distribution of the sparse signal through EM and thus achieves near MMSE result. Although the full understanding of EM algorithm is still not available, its efficiency can be proven empirically in this high SNR example. On the other hand, both P-GENP-AMP and P-GENP-AMP-FK perform almost the same as GENP-AMP with known GENP variance. The reason is that at high SNR, the MSE of AMP is very small. Therefore Eq. (\ref{var_true}) and Eq. (\ref{var_fake}) are very similar.

Fig. \ref{paraless_perf} (c) and (d) show the results with a low SNR of $5$ dB. In this case, EMGMAMP-GENP no longer achieves an accurate estimate of $\sigma _s^2$, whereas the proposed P-GEMP-AMP still performs well. Moreover, P-GENP-AMP and GENP-AMP are still very close and are much better than other algorithms. The failure of EMGMAMP-GENP is because there are many approximations in EMGMAMP, {\it e.g.}, using the GAMP approximated posterior as the true one and learning the hidden parameters through EM. At low SNRs, these approximations are not accurate, and the method cannot achieve near MMSE result. Its performance can be even worse than the AMP.

\begin{table*}[t]
\centering{
\vskip 10pt
\scalebox{0.75}{
\begin{tabular}{c*{10}{|c}}
\hline
Test sequence & ${\sigma ^2},\sigma _s^2$ & $\delta$  & Alg1 & Alg2 & Alg3 & Alg4 & Alg5 & Alg6 & Alg7 & Alg8  \\
\hline
\multirow{6}{*}{Balloons} & \multicolumn{1}{c|}{\multirow{2}{*}{$1{e2},1{e2}$}} & 1/5 & 31.27 & \textbf{33.72} & \textbf{33.72} & 32.65 & \textbf{34.50} & 27.25 & 32.04 & 32.31 \\
 & & 1/2 & 34.71 & \textbf{35.63} & \textbf{35.79} & 30.41 & 30.65 & 28.04 & 32.04 & 35.62 \\
\cline{2-11}
&  \multicolumn{1}{c|}{\multirow{2}{*}{$1{e2},1{e3}$}} & 1/5  & 31.27 & \textbf{32.71} & 32.61 & 32.65 & \textbf{33.20} & 18.02 & 28.69 & 14.28 \\
 & & 1/2 & 34.71 & \textbf{35.07} & \textbf{35.10} & 30.43 & 30.20 & 19.45 & 28.69 & 32.91 \\
\cline{2-11}
& \multicolumn{1}{c|}{\multirow{2}{*}{$1{e3},1{e3}$}} & 1/5 & 27.83 & \textbf{30.36} & \textbf{30.42} & 27.08 & 25.70 & 18.01 & 28.69 & 15.38 \\
& & 1/2 & 29.06 & \textbf{30.87} & \textbf{30.94} & 21.17 & 20.60 & 18.52 & 28.69 & 29.81 \\
\hline
\multirow{6}{*}{Kendo} & \multicolumn{1}{c|}{\multirow{3}{*}{$1{e2},1{e2}$}} & 1/5 & 33.08 & \textbf{35.88} & \textbf{35.82} & 34.37 & 35.56 & 27.57 & 33.51 & 34.77 \\
& & 1/2 & 36.22 & \textbf{37.05} & 37.04 & 30.79 & 30.89 & 28.28 & 33.51 & \textbf{37.33} \\
\cline{2-11}
& \multicolumn{1}{c|}{\multirow{3}{*}{$1{e2},1{e3}$}} & 1/5 & 33.08 & 34.73 & \textbf{34.76} & 34.37 & \textbf{35.20} & 18.07 & 30.20 & 16.77 \\
& & 1/2 & 36.22 & \textbf{36.63} & \textbf{36.64} & 30.77 & 30.59 & 19.50 & 30.20 & 35.11 \\
\cline{2-11}
& \multicolumn{1}{c|}{\multirow{3}{*}{$1{e3},1{e3}$}} & 1/5 & 28.15 & \textbf{31.86} & \textbf{32.00} & 28.07 & 25.98 & 18.04 & 30.20 & 22.30 \\
& & 1/2 & 30.26 & \textbf{32.20} & \textbf{32.31} & 21.32 & 20.64 & 18.57 & 30.20 & 31.04 \\
\hline
\multirow{6}{*}{Pantomime} & \multicolumn{1}{c|}{\multirow{3}{*}{$1{e2},1{e2}$}} & 1/5 & 31.65 & \textbf{34.41} & \textbf{34.20} & 33.42 & 33.51 & 27.43 & 31.93 & 24.79 \\
& & 1/2 & \textbf{36.46} & 36.24 & 36.36 & 30.89 & 30.29 & 28.20 & 31.93 & \textbf{37.62} \\
\cline{2-11}
& \multicolumn{1}{c|}{\multirow{3}{*}{$1{e2},1{e3}$}} & 1/5 & 31.65 & 33.73 & \textbf{33.77} & 33.42 & \textbf{34.40} & 18.06 & 29.77 & 24.58 \\
& & 1/2 & 36.46 & \textbf{36.62} & \textbf{36.66} & 30.88 & 30.57 & 19.48 & 29.77 & 34.41 \\
\cline{2-11}
& \multicolumn{1}{c|}{\multirow{3}{*}{$1{e3},1{e3}$}} & 1/5 & 28.50 & \textbf{31.39} & \textbf{31.49} & 28.01 & 25.74 & 17.63 & 29.77 & 26.38 \\
& & 1/2 & 30.32 & \textbf{31.86} & \textbf{32.01} & 21.34 & 20.66 & 18.56 & 29.77 & 31.11 \\
\hline
\end{tabular}
}
}
\vskip 5pt
\caption{\label{Testseq} PSNRs of different methods for multiview images. For $\sigma _s^2 = 1{e3}$, the PSNRs of the corrupted virtual middle views are all 18.03 dB, whereas when $\sigma _s^2 = 1{e2}$, the PSNRs are 26.96 \textnormal{d}B for "Balloons", 27.35 \textnormal{d}B for "Kendo", and 27.20 \textnormal{d}B for "Pantomime".}
\vskip -10pt
\end{table*}

\begin{figure*}[t]
\begin{center}
\begin{tabular}{@{\hspace{-1mm}}c@{\hspace{-1mm}}c@{\hspace{-1mm}}c@{\hspace{-1mm}}c@{\hspace{-1mm}}c@{\hspace{-1mm}}c@{\hspace{-1mm}}c@{\hspace{-1mm}}c}
    \includegraphics[width=1.4in]{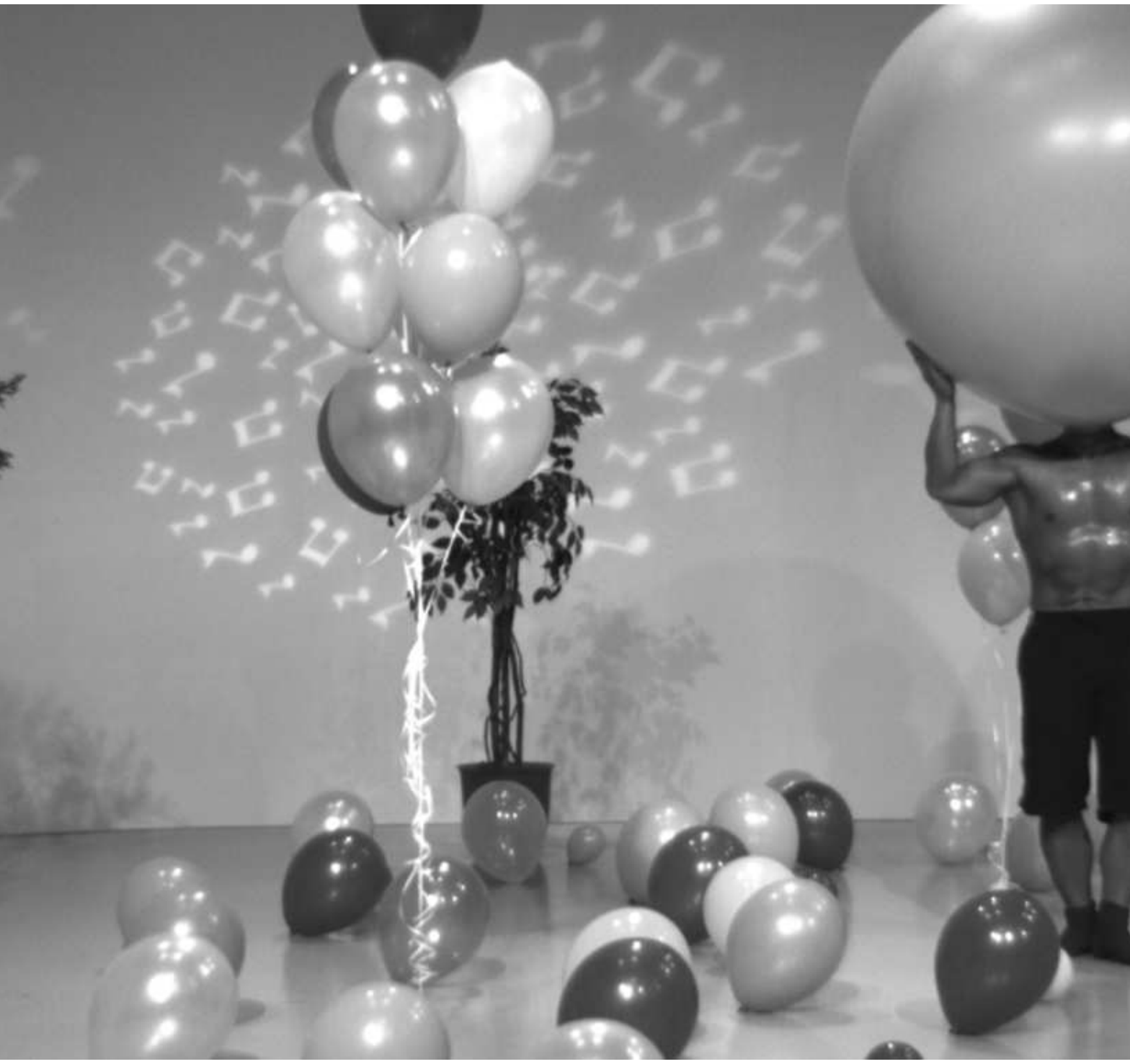} &
    \includegraphics[width=1.4in]{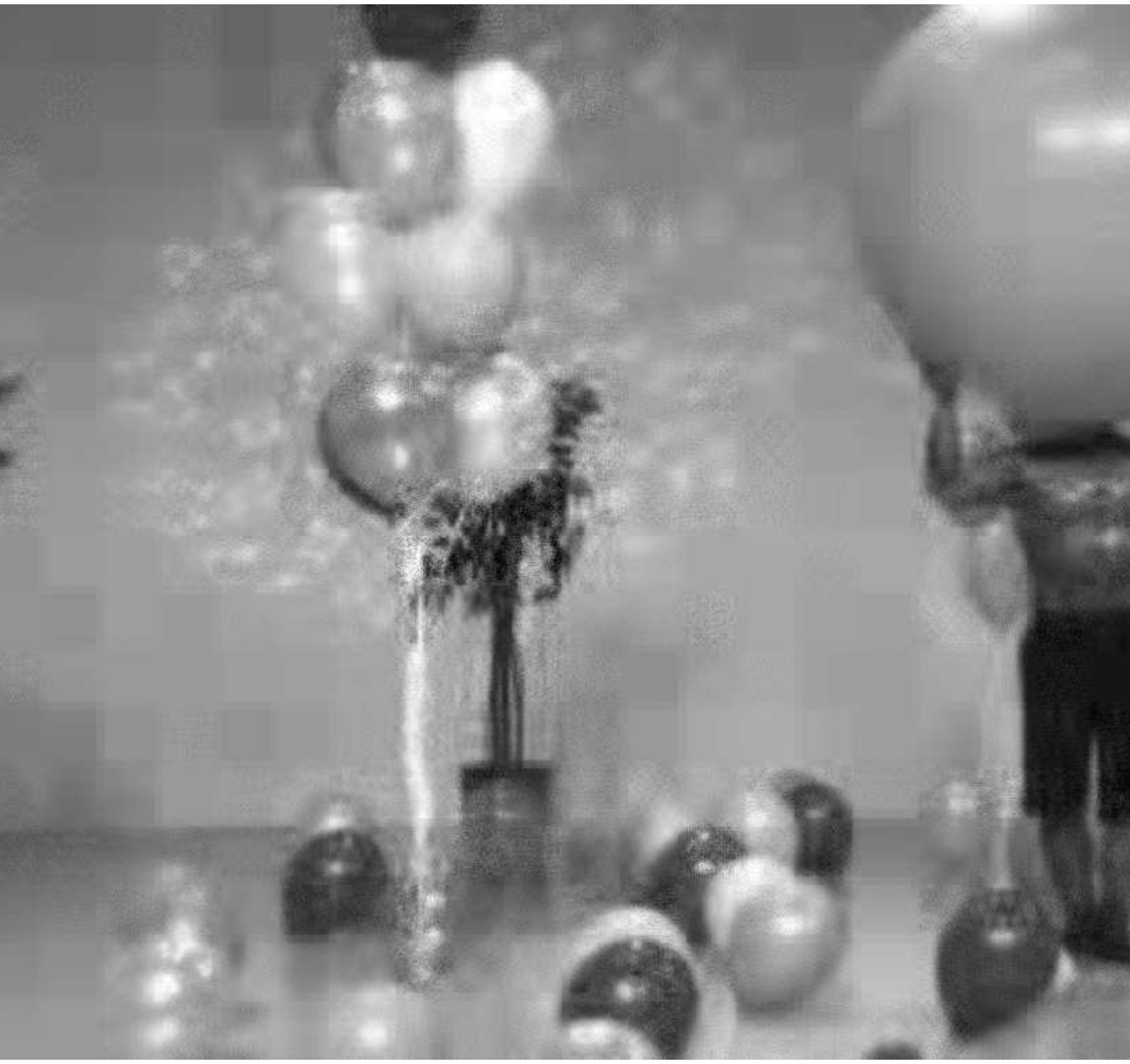} &
    \includegraphics[width=1.4in]{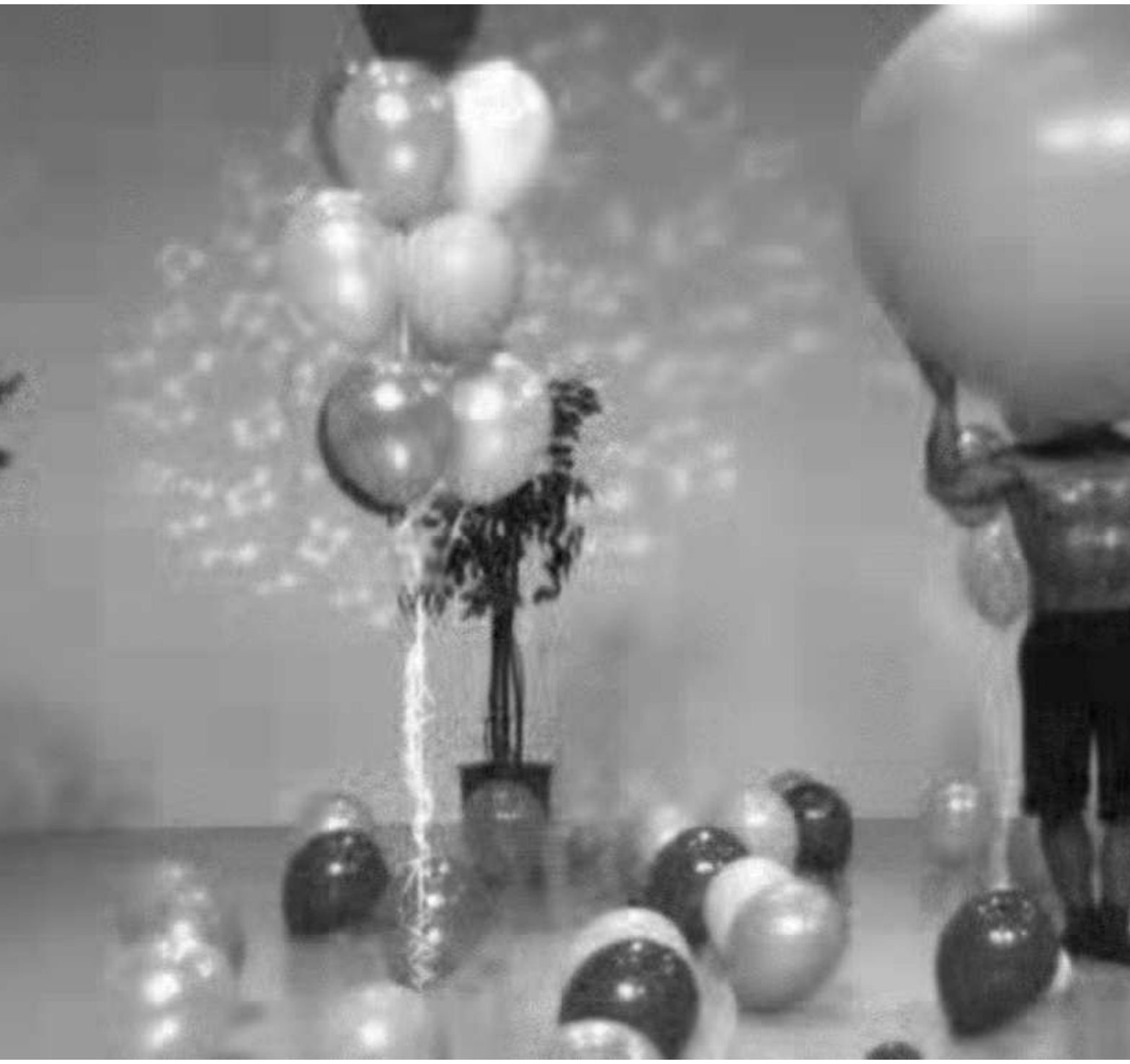} &
   \includegraphics[width=1.4in]{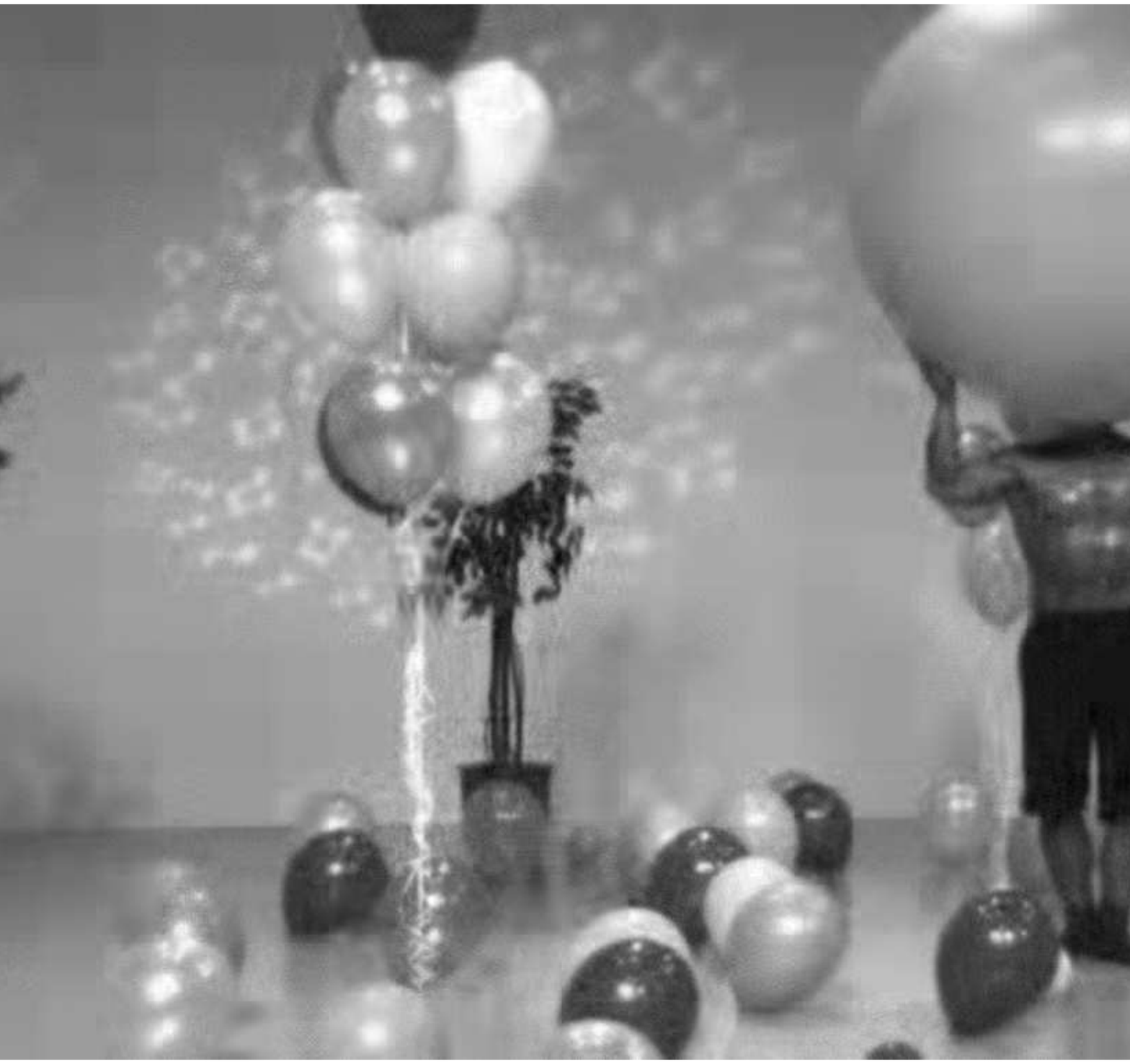} \\
    \includegraphics[width=1.4in]{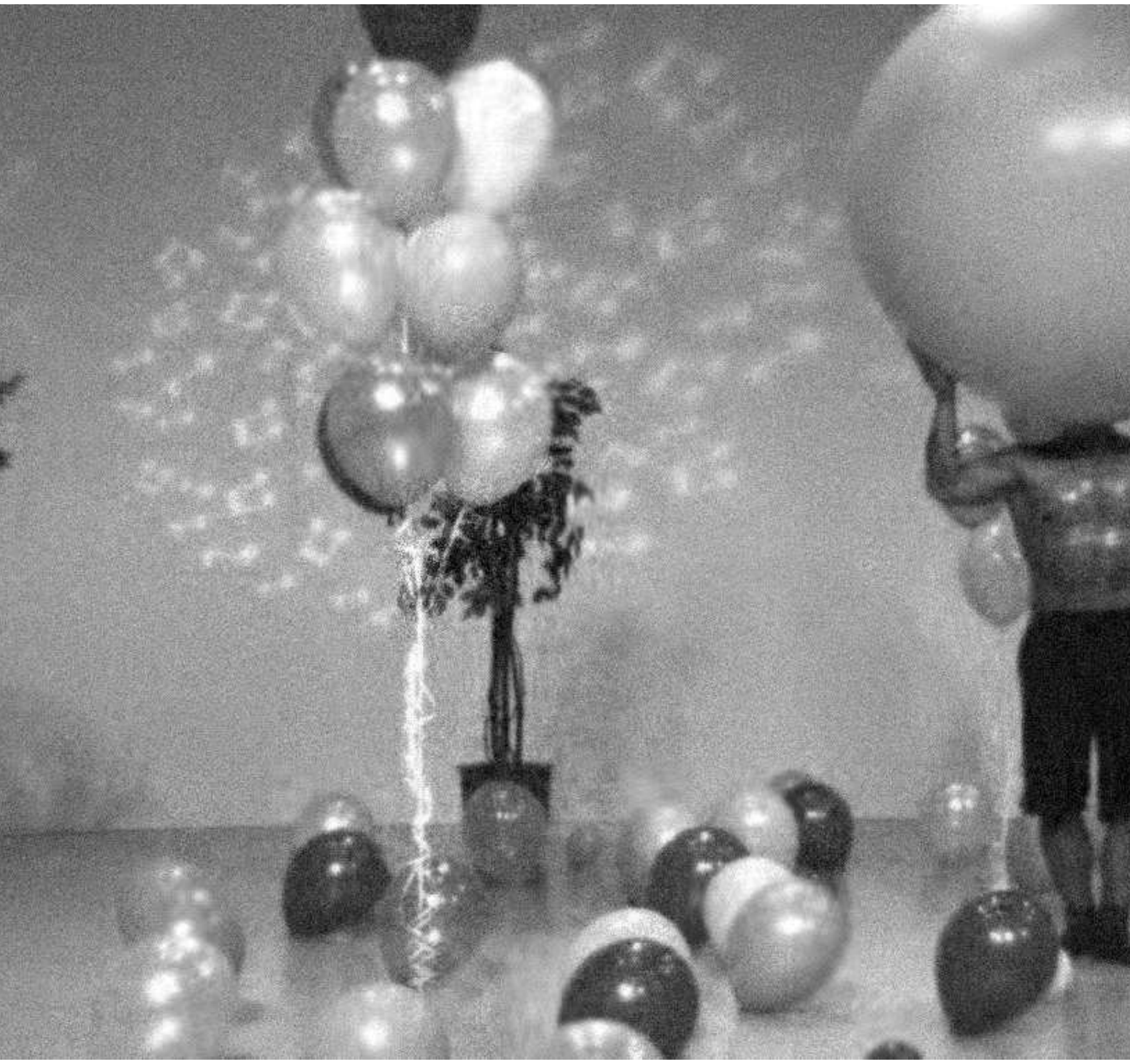} &
    \includegraphics[width=1.4in]{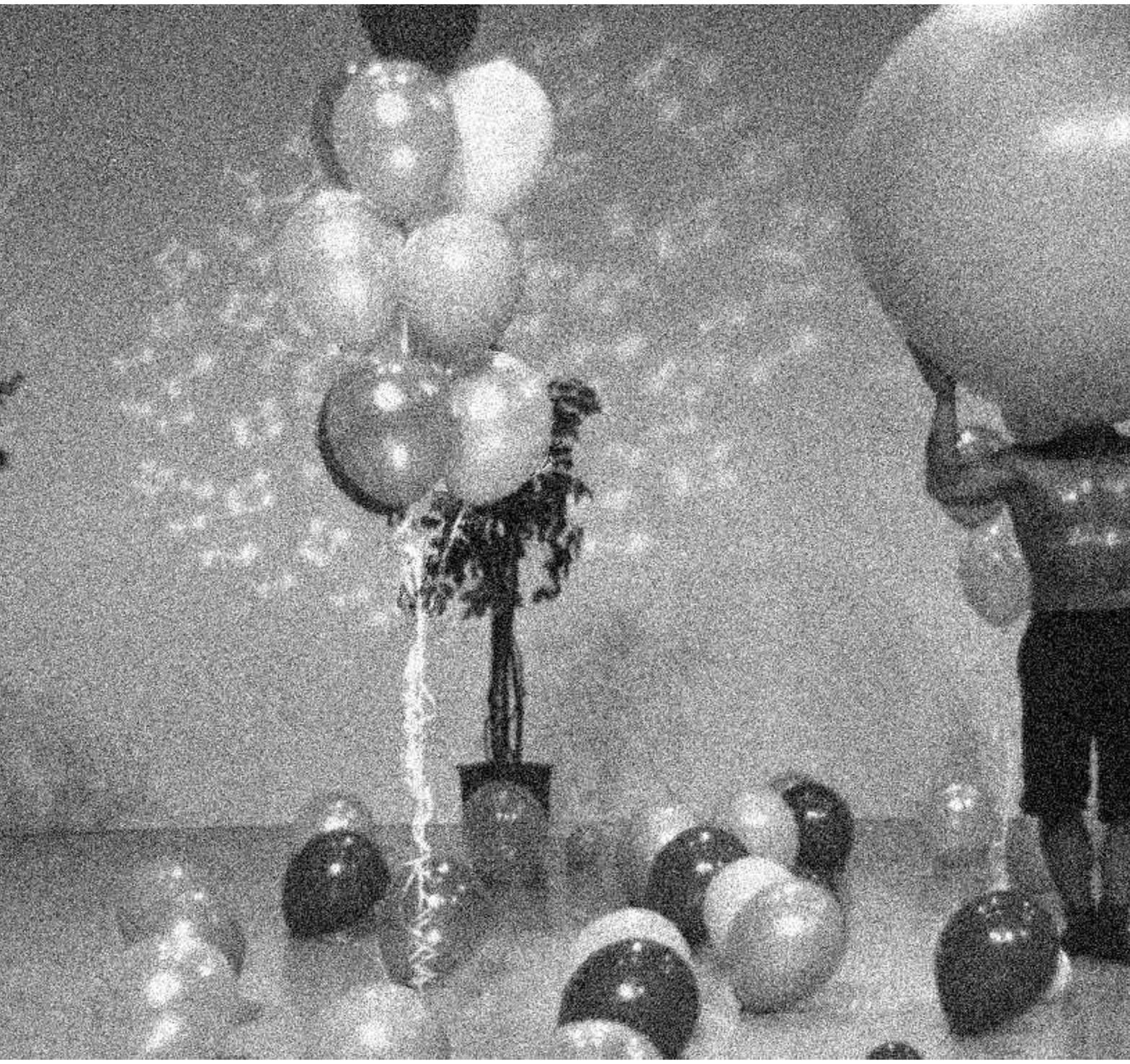} &
    \includegraphics[width=1.4in]{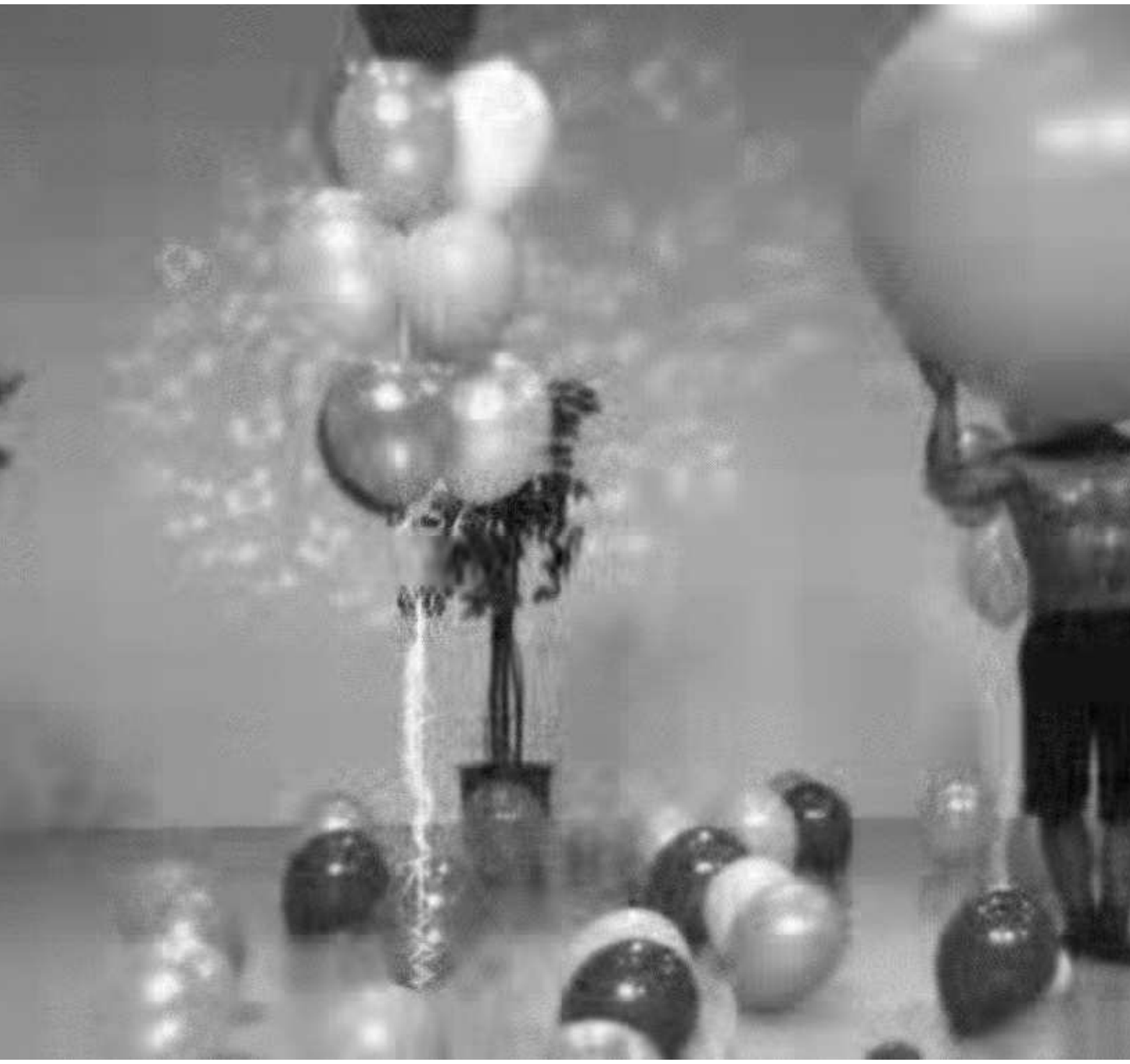} &
   \includegraphics[width=1.4in]{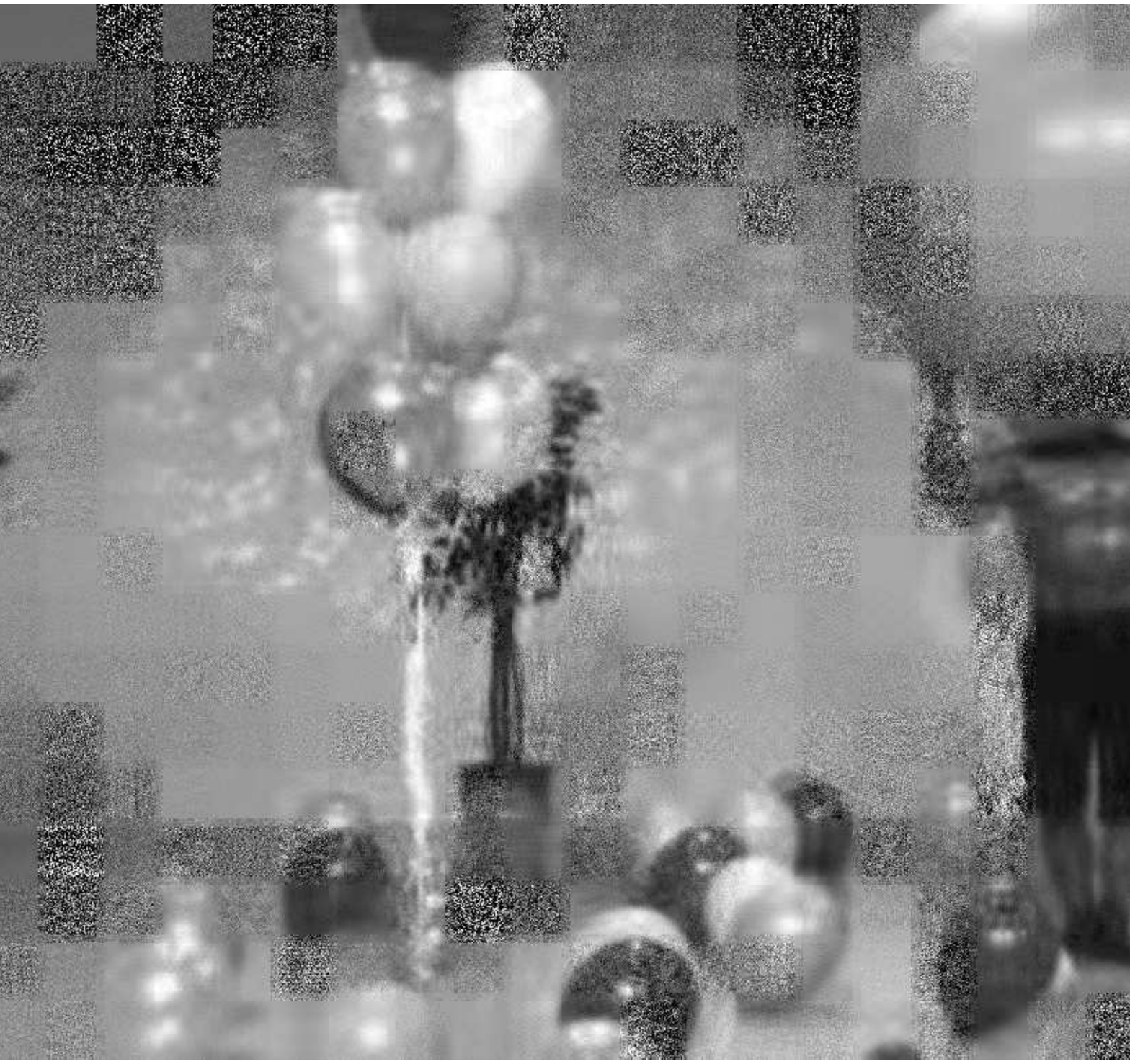} \\
\end{tabular}
\end{center}
\vskip -15pt
\caption{\label{BalloonsImage} The reconstructed "Balloons" with ${\sigma ^2} = 1{e3},\sigma _s^2 = 1{e3},\delta  = 1/5$. First row (from left to right): original, AMP (PSNR: 27.83dB) , P-GENP-AMP (30.36dB), GENP-AMP (30.42dB). Second row: EMGMAMP-GENP (25.70dB), Residual AMP (18.02dB), Denoising (28.69dB), and Modified CS (14.28dB).}
\vskip -10pt
\end{figure*}

\subsection{Application in Compressive Image Sampling}
\label{sec_image_exp}

We next consider a compressive image sensing example. The target image is the image "Lena" with resolution $512 \times 512$. We assume that the receiver has access to a $128 \times 128$ low-resolution version of the image, which is then upsampled to $512 \times 512$ and corrupted by Gaussian noises with different variances, to simulate the noises in poor illumination, high temperature, or transmission error. This is used as the GENP prior of our method.

The full size image is partitioned into overlapped blocks of size $48 \times 48$ pixels, with an overlap of 6 pixels to reduce the blocking artifacts. The DCT is used as the sparsifying transform. The same i.i.d. Gaussian sensing matrix is applied to each block to obtain the CS measurements. Eight algorithms are compared: AMP (denoted as Alg1), P-GENP-AMP (Alg2), GENP-AMP (Alg3), EMGMAMP (Alg4), EMGMAMP-GENP(Alg5), the residual AMP similar to \citep{Trocan01} (Alg6), the direct denoising of the prior image via soft-thresholding (Alg7), and the modified CS \citep{ModifiedCS} (Alg8), which finds the sparsest signal outside the support set detected from the prior $\tilde x$. For the denoising algorithm, the parameterless SURE framework in \citep{Paraless} is applied to automatically choose the tuning parameter, and $\sigma _s^2$ is assumed to be known.

The results are summarized in Table \ref{Upsampled}. The top-two best results in each case are highlighted in bold. We can see that our proposed P-GENP-AMP and GENP-AMP always outperform other algorithms. Besides, at low SNRs (${\sigma ^2} = 1e3$), the performance of EMGMAMP-GENP is quite poor. Note that the performance of Algorithms 4 and 5 degrade when given more samples, due to the instability of EM-based algorithms.

\subsection{Application in Hybrid Multi-View Imaging System}
\label{sec_MultiView}

We next apply the GENP-AMP to the hybrid multi-view imaging system \citep{Trocan01,Parmida,XingIcassp13}, where a group of cameras capture the scene from different locations. Some cameras are traditional cameras, and others are CS cameras such as the single pixel cameras \citep{singlePixel}. For each CS camera, we assume its left and right neighbouring cameras are traditional cameras. To help the reconstruction from CS sampling, the left and right views are used to generate a virtual view, which is corrupted by Gaussian noise and serves as the initial estimate or the GENP of the middle view.



We test the multiview image sequences "Balloons", "Kendo", and "Pantomime" under various channel noise levels. The setup is similar to Sec. \ref{sec_image_exp}. The virtual middle image is generated by Version 3.5 of the MPEG view synthesis reference software (VSRS) \citep{VSRS}, and the test sequences are downloaded from \citep{Nagoya}.

Table \ref{Testseq} reports the PSNRs (dB) of the reconstructions given by the eight methods under different $\sigma^2$, $\sigma_s^2$, and $\delta$. The following can be observed. First, almost all the top-two results are P-GENP-AMP and GENP-AMP, and there is no noticeable gap between them, verifying the efficiency of the proposed algorithms. In particular, when ${\sigma ^2} = 1e3$ and $\sigma _s^2 = 1e3$, {\it i.e.}, both the CS samples and GENP have low quality, our algorithms always perform the best. Second, when the channel noise level is low and sampling rate is high, {\it i.e.}, ${\sigma ^2} = 1{e2}$, $\sigma _s^2 = 1{e2}$, and $\delta  = 1/2$, the modified CS (Alg6) is comparable to or even better than the proposed methods Alg2 and Alg3. This is as expected, since detecting the support of the virtual view $\tilde x$ is easier under low noise levels. However, as the noise level increases, the performance of the modified CS degrades quickly. It also requires the knowledge of ${\sigma^2}$, which is not needed in AMP-based algorithms. Third, at high SNR (${\sigma ^2} = 1e2$), EMGMAMP-GENP outperforms the proposed P-GENP-AMP, but our method is better at low SNRs.  Finally, Our methods are also about 20 times faster than the CVX-based modified CS and comparable to EMGMAMP and EMGMAMP-GENP.

Some examples of the reconstructed images are shown in Fig. \ref{BalloonsImage}. Our P-GENP-AMP and GENP-AMP provide the best visual quality. All other methods have some limitations. For example, some artifacts exist in the AMP and EMGMAMP. Blurs happen when thresholding-based denoising is used, and Gaussian noises cannot be removed by the residual AMP. Although some parts can be well recovered by the modified CS, it also introduces severe artifacts in certain areas, due to its poor detection rate of the support set in high noise levels.

\section{Conclusions and Future Work}

This paper studies the generalized elastic net prior (GENP)-aided compressed sensing problem, where an additional noisy version of the original signal is available for CS reconstruction. We develop a GENP-aided approximate message passing algorithm (GENP-AMP), and study its parameter selection, state evolution, and noise sensitivity. The contribution of the GENP is also examined. We also develop a parameterless GENP-AMP that does not need to know the sparsity of the unknown signal and the variance of the GENP. Simulation results with 1-D data and two imaging applications demonstrate the performances of the proposed methods.

For the future work, a parameterless GENP-AMP algorithm that can accurately work in the whole plane need to be developed. According to the noise sensitivity analysis in Sec. \ref{noise_sensitivity}, there is no phase transition boundary, and the MSE is bounded in the whole plane. However, the parameterless GENP-AMP proposed in Sec. \ref{Parameterless} only works well below the phase transition boundary of the standard AMP, due to the unbounded MSE above the phase transition boundary of the standard AMP and the approximation accuracy of SURE.

The original AMP is based on the simple soft thresholding in each iteration. Recently, it is found in \citep{DAMP-NC,DAMP-Rice} that other denoising methods can be employed in AMP to further improve the reconstruction. For example, using the BM3D denoising algorithm \citep{BM3D}, state-of-the-art CS reconstructions can be achieved in imaging applications. This approach can also be adopted into the GENP-AMP framework in this paper.

Applying the proposed schemes to multiview videos instead of multiview images is another attractive topic, where the approaches in \citep{AMP-MMV,DCS-AMP} could be useful. It is also worthwhile to find other applications of the proposed GENP-AMP method.

\appendix

\section{A heuristic derivation of the state evolution of GENP-AMP}
\label{apx_state}

In this section, we derive the state evolution of GENP-AMP in Eq. (\ref{evolution}) of Sec. \ref{subsec_state}. The derivation is generalized from that in \citep{Graphical} for AMP. We start from the GENP-AMP iteration in (\ref{amp_x}) and (\ref{amp_r}), but introduce the following three modifications: (i) The random matrix $A$ is replaced by a new i.i.d. $A(t)$ at each iteration $t$, where ${A_{ij}}(t)\sim N(0,1/m)$; (ii) The corresponding observation becomes ${y^t} = A(t)x + w$; (iii) The last term in the update equation for ${r^t}$ is eliminated. We thus get the following dynamics:
\begin{equation}
{x^{t + 1}} = \eta (\frac{{{u_t}}}
{{1 + {u_t}}}\tilde x + \frac{1}
{{1 + {u_t}}}({x^t} + A{(t)^T}{r^t});{\theta _t}),
\end{equation}
\begin{equation}
{r^t} = {y^t} - A(t){x^t}.
\end{equation}

Eliminating ${r^t}$, the first equation becomes:
\begin{equation}
\label{derivation_state}
\small{
\begin{gathered}
  {x^{t + 1}} = \eta (\frac{{{u_t}}}
{{1 + {u_t}}}\tilde x + \frac{1}
{{1 + {u_t}}}(A{(t)^T}{y^t} + ({\text{I}} - A{(t)^T}A(t)){x^t};{\theta _t}) \hfill \\
  {\text{      }} = \eta (x + \frac{{{u_t}}}
{{1 + {u_t}}}(\tilde x - x) + \frac{1}
{{1 + {u_t}}}(A{(t)^T}w + B(t)({x^t} - x));{\theta _t}), \hfill \\
\end{gathered}
}
\end{equation}
where $B(t) = {\text{I}} - A{(t)^T}A(t)$.

Since the large system limit is assumed here, similar to \citep{AMP}, $q_t^2$ in  Sec. \ref{subsec_state} can be approximated by $\mathop {\lim }\limits_{n \to \infty } \left\| {{x^t} - x} \right\|_2^2/n$. It can be shown using the central limit theorem that $B(t)({x^t} - x)$ converges to a vector with i.i.d. normal entries, and each entry has zero mean and variance $q_t^2/\delta $. In addition, the entries of $A(t)^Tw$ have zero mean and variance of $\sigma^2$, and they are independent of $B(t)(x^t-x)$. Therefore, each entry of the vectors in the argument of $\eta $ in Eq. (\ref{derivation_state}) converges to ${X_0} + {\xi _t}Z$ with $Z\sim {\text{N}}(0,1)$ independent of ${X_0}$, and
\begin{equation}
\label{recursion1}
  \xi _t^2 = {\left( {\frac{{{u_t}}}
{{1 + {u_t}}}} \right)^2}\sigma_s^2 + {\left( {\frac{1}
{{1 + {u_t}}}} \right)^2}({\sigma ^2} + \frac{1}
{\delta }{q_t^2}). \hfill \\
\end{equation}

On the other hand, by Eq. (\ref{derivation_state}), each entry of ${x^{t + 1}} - x$ converges to $\eta ({X_0} + {\xi _t}Z;{\theta _t}) - {X_0}$. Therefore
\begin{equation}
\label{recursion2}
{q^2_{t + 1}} = \mathop {\lim }\limits_{n \to \infty } \frac{1}
{n}\left\| {{x^{t + 1}} - x} \right\|_2^2 = E\{ {[\eta ({X_0} + {\xi _t}Z;{\theta _t}) - {X_0}]^2}\}.
\end{equation}

From Eq. (\ref{recursion1}) and Eq. (\ref{recursion2}), we can obtain the state evolution in Eq. (\ref{evolution}).

This is a heuristic proof, more rigorous proof can be achieved following the proof in \citep{dynamicsofAMP}.

\section{Proof of Proposition \ref{phase transition}}
\label{apx_phase}

In this part, we prove Prop. \ref{phase transition}, which studies the bound of the MSE of the GENP-AMP in the $(\rho,\delta)$ plane.

\begin{proof}
Consider ${p_0} \in {\mathcal{F}_{\delta \rho }}$, ${\sigma ^2} = 1$ and let ${\alpha ^*}(\delta ,\rho ) = {\alpha ^ \pm }(\delta \rho )$ minimax the MSE. To simplify the notation, we define
\beq
\begin{split}
\Psi ({q^2},u;p) &= \Psi ({q^2},u,\delta ,\sigma  = 1,{\sigma _s},{\alpha ^*},p) \\
&= mse(npi({q^2},u,1,{\sigma _s},\delta );p,{\alpha ^*}).
\end{split}
\eeq
Then, by the definition of fixed point, we get
\begin{equation}\nonumber
\begin{gathered}
  q_*^2 = \Psi (q_*^2,{u^*};p ), \hfill \\
  {u^*} = \frac{{1 + \frac{{q_*^2}}
{\delta }}}
{{\gamma_s^2}}. \hfill \\
\end{gathered}
\end{equation}

Using the scale invariance, we have $mse({\sigma ^2};p,{\alpha ^*}) = {\sigma ^2}mse(1;\tilde p,{\alpha ^*})$, where $\tilde p$ is a rescaled probability measure, $\tilde p\{ x \cdot \sigma  \in B\}  = p\{ x \in B\} $. For $p \in {F_{\delta \rho }}$, we have $\tilde p \in {F_{\delta \rho }}$ as well. Therefore,
\begin{equation}\nonumber
\begin{gathered}
  q_*^2 = mse(npi(q_*^2,{u^*},1,{\sigma_s},\delta );p,{\alpha ^*}) \hfill \\
  \quad = mse(1;\tilde p,{\alpha ^*}) \cdot npi(q_*^2,{u^*},1,{\sigma_s},\delta ) \hfill \\
  \quad \leqslant {M^ \pm }(\delta \rho )\cdot npi(q_*^2,{u^*},1,{\sigma_s},\delta ) \hfill \\
\end{gathered}
\end{equation}

Hence,
\begin{equation}\nonumber
\frac{{q_*^2}}
{{npi(q_*^2,{u^*};1,{\sigma _s},\delta )}} \leqslant {M^ \pm }(\delta \rho ),
\end{equation}
where we use the fact that $\sigma  = 1$ and ${\gamma _s} = {\sigma _s}$.

By the definition of npi in Eq. (\ref{npi}), we have
\begin{equation}\nonumber
\frac{{q_*^2}}
{{{{(\frac{{{u^*}}}
{{1 + {u^*}}})}^2}\gamma_s^2 + {{(\frac{1}
{{1 + {u^*}}})}^2}(1 + \frac{{q_*^2}}
{\delta })}} \leqslant {M^ \pm }(\delta \rho ).
\end{equation}

Replacing ${u^*}$ by (\ref{weight}), we get
\begin{equation}
\label{minimax_inequ}
q_*^2 \leqslant \frac{{ - G(\delta ,\rho ,\gamma _s^2) + \sqrt {G{{(\delta ,\rho ,\gamma _s^2)}^2} + 4\delta \gamma _s^2{M^ \pm }(\delta \rho )} }}{2}
\end{equation}
where $G(\delta ,\rho ,\gamma _s^2) = \delta \gamma _s^2 + \delta  - \gamma _s^2{M^ \pm }(\delta \rho )$.

It is easy to verify that the phase transition boundary only exists when $\gamma _s^2 = \infty $ from the inequality above.
If we let $(\gamma_s^2 + 1)\delta  < \gamma_s^2{M^\pm }(\delta \rho )$, $G(\delta ,\rho ,\gamma _s^2)$
in the right hand side of Eq. (\ref{minimax_inequ}) is positive. In such case, if $\gamma_s^2$ goes to $\infty $, then $\delta  < {M^ \pm }(\delta \rho )$, we can get $q_*^2 \leqslant \infty $, i.e., the mean square error is unbounded, corresponding to the classical AMP phase transition boundary.

To prove the second part of Prop. \ref{phase transition}, we make a specific choice $\bar p$ of $p$, and fix a small constant $c > 0$.

Now for $\varepsilon  = \delta \rho $, define $h = {h^ \pm }(\varepsilon ,c){\text{ }} \cdot {\text{ }}\sqrt {{\text{NP}}{{\text{I}}^*}} $. Let $\bar p = (1 - \varepsilon ){\delta _0} + (\varepsilon /2){\delta _{ - h}} + (\varepsilon /2){\delta _h}$, similar to (\ref{least_favor_p}). Denote $q_*^2 = q_*^2(\bar p)$ the highest fixed point corresponding to the signal distribution. Again, by the scale invariance, we have
\begin{equation}\nonumber
\begin{split}
q_*^2 &= mse(npi(q_*^2,{u^*},1,{\gamma_s},\delta );\bar p,{\alpha ^*}) \\
&= mse(1;\tilde p,{\alpha ^*}) \cdot npi(q_*^2,1,{\gamma_s},\delta ),
\end{split}
\end{equation}
where $\tilde p$ is a scaled probability measure, and $\tilde p\{ x \cdot \sqrt {npi(q_*^2,1,{\gamma_s},\delta )}  \in B\}  = \bar p\{ x \in B\} $. Since $q_*^2 \leqslant {M^*}$, we have ${\text{npi}}(q_*^2,1,{\gamma_s},\delta ) \leqslant {\text{NP}}{{\text{I}}^*}$ and hence
\begin{equation}\nonumber
\frac{h}
{{\sqrt {{\text{npi(}}q_*^2,1,{\gamma_s},\delta )} }} = {h^ \pm }(\varepsilon ,c){\text{ }} \cdot {\text{ }}\sqrt {\frac{{{\text{NP}}{{\text{I}}^*}}}
{{{\text{npi}}(q_*^2,1,{\gamma_s},\delta )}}}  > {h^ \pm }(\varepsilon ,c).
\end{equation}

Note that ${\text{mse(}}q;(1 - \varepsilon ){\delta _0} + (\varepsilon /2){\delta _{ - x}} + (\varepsilon /2){\delta _x},\alpha )$ increases monotonically in $\left| x \right|$. Recall that ${p_{\varepsilon ,c}} = (1 - \varepsilon ){\delta _0} + (\varepsilon /2){\delta _{ - {h^ \pm }(\varepsilon ,c)}} + (\varepsilon /2){\delta _{{h^ \pm }(\varepsilon ,c)}}$ is nearly-least-favorable for the minimax problem. Consequently,
\begin{equation}\nonumber
{\text{mse}}(1;\tilde p,{\alpha ^*}) \geqslant {\text{mse}}(1;{p_{\delta \rho ,c}},{\alpha ^*}) = (1 - c){\text{ }} \cdot {\text{ }}{M^ \pm }(\delta ,\rho ).
\end{equation}

By the scale-invariant property, we conclude that
\begin{equation}\nonumber
\frac{{q_*^2}}
{{{\text{npi}}(q_*^2,1,{\gamma_s},\delta )}} \geqslant (1 - c){\text{ }} \cdot {\text{ }}{M^ \pm }(\delta \rho ).
\end{equation}

Then, we can get the inequality
\begin{equation}\nonumber
\begin{split}
{(q_*^2)^2} + [\delta (\gamma_s^2 + 1) &- (1 - c){M^ \pm }(\delta ,\rho )\gamma_s^2]q_*^2 \\
&- (1 - c){M^ \pm }(\delta \rho )\gamma_s^2\delta  \geqslant 0.
\end{split}
\end{equation}
Therefore,
\begin{equation}\nonumber
\small{
\begin{split}
&{\text{fMSE}}({\alpha ^*};\delta ,\rho ,1,\gamma_s^2,\bar p)
\geqslant \frac{-[\delta (\gamma_s^2 + 1) - (1 - c){M^ \pm }(\delta ,\rho )\gamma_s^2]}{2} \\
&+ \frac{\sqrt {{{[\delta (\gamma_s^2 + 1) - (1 - c){M^ \pm }(\delta ,\rho )\gamma_s^2]}^2} + 4(1 - c){M^ \pm }(\delta \rho )\gamma_s^2\delta }}{2},
\end{split}
}
\end{equation}
where ${\text{fMSE}}(\alpha ;\delta ,\rho ,\sigma ,\gamma_s^2,p)$ is the equilibrium formal MSE for GENP-AMP ($\lambda$, $\tau_s$) for the large system framework \citep{AMP}.

As $c > 0$ is arbitrary, we conclude
\begin{equation}\nonumber
\begin{split}
&\mathop {\sup }\limits_{p \in {F_{\delta \rho }}} {\text{fMSE}}({\alpha ^*};\delta ,\rho ,1,\gamma_s^2,p) \geqslant
\frac{-[\delta (\gamma_s^2 + 1) - {M^ \pm }(\delta ,\rho )\gamma_s^2]}{2}\\
&+ \frac{\sqrt {{[\delta (\gamma_s^2 + 1) - {M^ \pm }(\delta ,\rho )\gamma_s^2]}^2 + 4{M^ \pm }(\delta \rho )\gamma_s^2\delta}} {2}.
\end{split}
\end{equation}

Also, following the same procedure as Prop. 4.2 in \citep{AMP}, it can be shown that ${M^*} = \mathop {\inf }\limits_\alpha  \mathop {\sup }\limits_{p \in {F_{\delta \rho }}} {\text{fMSE(}}\alpha {\text{;}}\delta {\text{,}}\rho {\text{,}}\sigma {\text{ = 1,}}\gamma_s^2,p)$.

The last part of Prop. \ref{phase transition} can be proven by simply substituting the fixed point results in the second part of Prop. \ref{phase transition} for the ones in Eq. (\ref{parameter_prediction}).
\end{proof}

\section{Proof of Proposition \ref{gamma_appro}}
\label{gamma_proof}

In this part, we prove Prop. \ref{gamma_appro}, which provides an accurate estimation of the variance of the prior $\tilde x$, {\it i.e.}, $\sigma _s^2$. This is an important step of the parameterless GENP-AMP.

\begin{proof}
From the definition of the GENP $\tilde x$, we get
\begin{equation}
\label{gamma_part}
\begin{gathered}
  \sigma _s^2 = E[{(\tilde X - {X_0})^2}]{\text{ }} \hfill \\
   = E[{(\tilde X - {X_{{\text{pos}}}} - {X_0} + {X_{{\text{pos}}}})^2}] \hfill \\
   = \underbrace {E[{{(\tilde X - {X_{{\text{pos}}}})}^2}]}_{(a)} + \underbrace {E[{{({X_0} - {X_{{\text{pos}}}})}^2}]}_{(b)} \hfill \\
  {\text{   }} - 2\underbrace {E[(\tilde X - {X_{{\text{pos}}}})({X_0} - {X_{{\text{pos}}}})]}_{(c)} \hfill \\
\end{gathered}
\end{equation}
where ${X_{{\text{pos}}}}$ is the estimated sparse signal by GENP-AMP based on an postulated variance $\sigma _{{\text{s-pos}}}^2$. $\tilde X$ and ${X_{{\text{pos}}}}$ can be explicitly expressed as follows.
\begin{equation}
\begin{gathered}
  \tilde X = {X_0} + e,{\text{          }}e \sim N(0,\sigma _s^2) \hfill \\
  {X_{{\text{pos}}}} = \eta ({X_0} + {\sigma _*}Z;\theta ),{\text{     }}Z \sim N(0,1) \hfill,
\end{gathered}
\end{equation}
where $\sigma _*^2$ is the variance of the unthresholded estimator in the last iteration of GENP-AMP.

Next, we look at each part of Eq. (\ref{gamma_part}). Part (c) can be rewritten as
\begin{equation}
E[(\tilde X - {X_{{\text{pos}}}})({X_0} - {X_{{\text{pos}}}})] = E[{({X_0} - {X_{{\text{pos}}}})^2}] + E[e({X_0} - {X_{{\text{pos}}}})].
\end{equation}
Thus Eq. (\ref{gamma_part}) becomes
\begin{equation}
\label{appro_trans}
\sigma _s^2 = E[{(\tilde X - {X_{{\text{pos}}}})^2}] - E[{({X_0} - {X_{{\text{pos}}}})^2}] - 2E[e({X_0}-{X_{{\text{pos}}}})].
\end{equation}

If $\sigma _{{\text{s-pos}}}^2$ is set to $\infty $, GENP-AMP degrades to AMP, which does not use $\tilde x$. This implies that a perfect candidate of ${X_{{\text{pos}}}}$ is the signal recovered by AMP, ${X_{{\text{AMP}}}}$.
Therefore, the two Gaussian noises  ${\sigma _*}Z$ and $e$ are uncorrelated. As a result, $E[e({X_0} - {X_{{\text{AMP}}}})] = 0$, and $\sigma _s^2$ can be further represented as
\begin{equation}
\sigma _s^2 = E[{(\tilde X - {X_{{\text{AMP}}}})^2}] - E[{({X_0} - {X_{{\text{AMP}}}})^2}].
\end{equation}

Part (a) can be rewritten as $E[{(\overset{\hbox{$\smash{\scriptscriptstyle\frown}$}}{X}  - \eta (\overset{\hbox{$\smash{\scriptscriptstyle\frown}$}}{X}  + {\sigma _*}Z - e;\theta ))^2}]$
This term can exactly be seen as a denoising operator. According to the large system limit \citep{AMP}, when $n$ is sufficiently large,
\begin{equation}
E[{(\tilde X - {x_{{\text{AMP}}}})^2}] \approx \frac{{\left\| {\tilde x - {x_{{\text{AMP}}}}} \right\|_2^2}}
{n}.
\end{equation}

Next, $E[{({X_0} - {X_{{\text{AMP}}}})^2}]$ can be estimated by the method proposed in \citep{Paraless}, inspired by the SURE theory. According to Theorem 4.3 and Theorem 4.7 in \citep{Paraless}, it can be predicted by $\mathop {\lim }\limits_{N \to \infty } \frac{{{{\overset{\hbox{$\smash{\scriptscriptstyle\frown}$}}{r} }^t}({\tau ^t})}}
{N}$ when  $t \to \infty $,  where $t$ is the inner iteration index of AMP. Usually it will converge in a few iterations.

Summarizing the analyses above, we can prove Prop. \ref{gamma_appro}.
\end{proof}

\section*{Acknowledgement}
{
The authors thank the reviewers for their suggestions that have significantly enhanced the quality and presentation of the paper.
}

\section*{References}

\bibliographystyle{elsarticle-num}
\bibliography{refs}

\begin{thebibliography}{10}
\expandafter\ifx\csname url\endcsname\relax
  \def\url#1{\texttt{#1}}\fi
\expandafter\ifx\csname urlprefix\endcsname\relax\def\urlprefix{URL }\fi
\expandafter\ifx\csname href\endcsname\relax
  \def\href#1#2{#2} \def\path#1{#1}\fi

\bibitem{RIP}
E.~J. Cand\`{e}s, T.~Tao, Decoding by linear programming, IEEE Transaction on
  Information Theory 51~(12) (2005) 4203--4215.

\bibitem{Coherence}
J.~A. Tropp, A.~C. Gilbert, Signal recovery from random measurements via
  orthogonal matching pursuit, IEEE Transaction on Information Theory 53~(12)
  (2007) 4655--4666.

\bibitem{IST}
A.~Beck, M.~Teboulle, A fast iterative shrinkage-thresholding algorithm for
  linear inverse problems, SIAM Journal in Imgace Sciences 2~(1) (2009)
  183--202.

\bibitem{Replica}
S.~Rangan, A.~K. Fletcher, V.~K. Goyal, Asymptotic analysis of {MAP} estimation
  via the replica method and applications to compressed sensing, IEEE
  Transaction on Information Theory 58~(3) (2012) 1902--1923.

\bibitem{lasso}
R.~Tibshirani, Regression shrinkage and selection with the lasso, J. Royal.
  Statist. Soc. B 58 (1996) 267--288.

\bibitem{BPDN}
S.~S. Chen, D.~L. Donoho, M.~A. Saunders, Atomic decomposition by basis
  pursuit, SIAM Journal on Scientific Computing 20~(1) (1998) 33--61.

\bibitem{AMPoriginal}
D.~Donoho, A.~Maleki, A.~Montanari, Message passing algorithms for compressed
  sensing, Proceedings of the National Academy of Sciences 106~(45) (2009)
  18914--18919.

\bibitem{AMP}
D.~Donoho, A.~Maleki, A.~Montanari, The noise-sensitivity phase transition in
  compressed sensing, IEEE Transaction on Information Theory 57~(10) (2011)
  6920--6941.

\bibitem{Graphical}
A.~Montanari, Graphical models concepts in compressed sensing, in: Compressed
  Sensing Theory and Applications, Cambrige University Press, 2012, pp.
  394--438.

\bibitem{AMPthesis}
A.~Maleki, Approximate message passing algorithms for compressed sensing, Ph.D.
  thesis, Stanford University (2010).

\bibitem{Sumproduct}
F.~R. Kschischang, B.~J. Frey, H.~A. Loeliger, Factor graphs and the
  sum-product algorithm, IEEE Transaction on Information Theory 47~(2) (2001)
  498--519.

\bibitem{GAMP}
S.~Rangan, Generalized approximate message passing for estimation with random
  linear mixing, arXiv: 1010.5141.

\bibitem{EMAMP}
J.~P. Vila, P.~Schniter, Expectation-{M}aximization {G}aussian-mixture
  approximate message passing, IEEE Transaction on Signal Processing 61~(19)
  (2013) 4658--4672.

\bibitem{Vila132}
J.~P. Vila, P.~Schniter, An empirical-{B}ayes approach to recovering linearly
  constrained non-negative sparse signals, ar{Xiv}: 1310.2806.

\bibitem{StructuredAMP}
S.~Som, L.~C. Potter, P.~Schniter, On approximate message passing for
  reconstruction of non-uniformly sparse signals, in: IEEE National Aerospace
  and Electronics Conference, 2010, pp. 223--229.

\bibitem{AMP-MMV}
J.~Ziniel, P.~Schniter, Efficient high-dimensional inference in the multiple
  measurement vector problem, IEEE Transactions on Signal Processing 61~(2)
  (2013) 340--354.

\bibitem{DCS-AMP}
J.~Ziniel, P.~Schniter, Dynamic compressive sensing of time-varying signals via
  approximate message passing, IEEE Transactions on Signal Processing 61~(21)
  (2013) 5270--5284.

\bibitem{XiongDSC}
Z.~Xiong, A.~D. Liveris, S.~Cheng, Distributed source coding for sensor
  networks, IEEE Signal Process. Mag. (2004) 80--94.

\bibitem{Trocan01}
M.~Trocan, T.~Maugey, J.~E. Fowler, B.~Pesquet-Popescu, Disparity-compensation
  compressed-sensing reconstruction for multiview images, in: IEEE
  International Conference on Multimedia and Expo, 2010, pp. 1225--1228.

\bibitem{singlePixel}
M.~F. Duarte, M.~A. Davenport, D.~Takhar, J.~N. Laska, T.~Sun, K.~F. Kelly,
  R.~G. Baraniuk, Single-pixel imaging via compressive sampling, IEEE Signal
  Processing Magazine 25~(2) (2008) 83--91.

\bibitem{LiWei}
L.~W. Kang, C.~S. Lu, Distributed compressive video sensing, in: IEEE
  International Conference on Acoustics, Speech and Signal Processing, 2009,
  pp. 1169--1172.

\bibitem{Parmida}
P.~Beigi, X.~Xiu, J.~Liang, Compressive sensing based multiview image coding
  with belief propagation, in: Proc. Asilomar Conference on Signals, Systems,
  and Computers, 2010, pp. 430--433.

\bibitem{XingIcassp13}
X.~Wang, J.~Liang, View interpolation confidence-aided compressed sensing of
  multiview images, in: IEEE International Conference on Acoustics, Speech, and
  Signal Processing, 2013, pp. 1651--1655.

\bibitem{DynamicCS}
A.~S. Charles, M.~S. Asif, J.~Romberg, C.~J. Rozell, Sparsity penalties in
  dynamic system estimation, in: Conference on Information Science and Systems,
  2011, pp. 1--6.

\bibitem{ModifiedCS}
N.~Vaswani, W.~Lu, Modified-{CS}: Modifying compressive sensing for problems
  with partially known support, IEEE Transaction on Signal Processing 58~(9)
  (2010) 4595--4607.

\bibitem{Hui}
H.~Zou, T.~Hastie, Regularization and variable selection via elastric net,
  Journal of the Royal Statistical Society: Series B (Statistical Methodology)
  67~(2) (2005) 301--320.

\bibitem{SZ13}
J.~Ziniel, P.~Schniter, Binary linear classification and feature selection via
  generalized approximate message passing, arXiv: 1401.0872.

\bibitem{Paraless}
A.~Mousavi, A.~Maleki, R.~G. Baraniuk, Parameterlss optimal approximate message
  passing, arXiv: 1311.0035.

\bibitem{CSBP}
R.~Baron, S.~Sarvoham, R.~G. Baraniuk, Bayesian compressive sensing via belief
  propagation, IEEE Trans. Signal Proc. 58~(1) (2010) 269--280.

\bibitem{GPSR}
M.~A.~T. Figueiredo, R.~D. Nowak, S.~J. Wright, Gradient projection for sparse
  reconstruction, IEEE Journal of Selected Topics in Signal Processing 1~(4)
  (2007) 586--597.

\bibitem{Kamilov}
U.~S. Kamilov, S.~Rangan, A.~K. Fletcher, M.~Unser, Approximate message passing
  with consistent parameter estimation and applications to sparse learning,
  IEEE Trans. Inf. Theory 60~(5) (2014) 2969--2985.

\bibitem{XingIcassp14}
X.~Wang, J.~Liang, Side information-aided compressed sensing reconstruction via
  approximate message passing, in: IEEE International Conference on Acoustics,
  Speech, and Signal Processing, 2014, pp. 3354--3358.

\bibitem{Mota14}
J.~Mota, N.~Deligiannis, M.~Rodrigues, Compressed sensing with prior
  information: Optimal strategies, geometry, and bounds, submitted to IEEE
  Trans. Info. Theory, arXiv: 1408.5250.

\bibitem{Renna14}
F.~Renna, L.~Wang, X.~Yuan, J.~Yang, G.~Reeves, R.~Calderbank, L.~Carin,
  M.~R.~D. Rodrigues, Classification and reconstruction of high-dimensional
  signals from low-dimensional noisy features in the presence of side
  information, preprint, arXiv: 1412.0614.

\bibitem{softshrink}
D.~L. Donoho, I.~M. Johnstone, Ideal spatial adaptation via wavelet shrinkage,
  Biomefrika 81~(3) (1994) 425--455.

\bibitem{BM3D}
K.~Dabov, A.~Foi, V.~Katkovnik, K.~Egiazarian, Image denoising by sparse 3-{D}
  transform-domain collaborative filtering, IEEE Trans. Image Proc. 16~(8)
  (2007) 2080--2095.

\bibitem{cvx}
M.~Grant, S.~Boyd, {CVX}: Matlab software for disciplined convex programming,
  version 2.0 beta, \url{http://cvxr.com/cvx} (Sep. 2013).

\bibitem{OWLQN}
G.~Andrew, J.~Gao, Scalable training of ${\ell_1}$-regularized log-linear
  models, in: Proc. of International Conference on Machine Learning, 2007, pp.
  33--40.

\bibitem{gamptoolbox}
S.~Rangan, A.~Fletcher, V.~Goyal, U.~Kamilov, J.~Parker, P.~Schniter, J.~Vila,
  J.~Ziniel, M.~Borgerding, {gampmatlab}: Generalized approximate message
  passing, \url{http://sourceforge.net/projects/gampmatlab/files/} (May. 2014).

\bibitem{VSRS}
M.~Tanimoto, T.~Fujii, K.~Suzuki, View synthesis algorithm in view synthesis
  reference software 3.5 document m16090, ISO/IEC JTC1/SC29/WG11 (MPEG).

\bibitem{Nagoya}
Fujii lab's multi-view sequences download lists,
  \url{http://www.fujii.nuee.nagoya-u.ac.jp/multiview-data/}.

\bibitem{DAMP-NC}
J.~Tan, Y.~Ma, D.~Baron, Compressive imaging via approximate message passing
  with image denoising, preprint, arXiv: 1405.4429.

\bibitem{DAMP-Rice}
C.~A. Metzler, A.~Maleki, R.~G. Baraniuk, From denoising to compressed sensing,
  preprint, arXiv: 1406.4175.

\bibitem{dynamicsofAMP}
M.~Bayati, A.~Montanari, The dynamics of message passing on dense graphs, with
  applications to compressed sening, IEEE Trans. Inf. Theory 57~(2) (2011)
  1462--1474.

\end{thebibliography}


\end{document}